\documentclass[useAMS,usenatbib,usegraphicx]{mn2e}
\usepackage{amsmath}
\usepackage{amssymb}
\usepackage{booktabs}
\usepackage{times}
\usepackage{aas_macros} 
\usepackage[usenames]{color}

\title[Main-sequence gravity mode pulsators]{A unifying explanation of complex frequency spectra of $\mbox{\boldmath$\gamma$}$~Dor, SPB and Be stars: combination frequencies and highly non-sinusoidal light curves
}

\author[Kurtz et al.]
{Donald W. Kurtz$^1$, Hiromoto Shibahashi$^2$, Simon J. Murphy$^{3,4}$, Timothy R. Bedding$^{3,4}$,
\newauthor{Dominic M. Bowman$^1$} \\
$^{1}$Jeremiah Horrocks Institute, University of Central Lancashire, Preston PR1 2HE, UK\\
$^{2}$Department of Astronomy, School of Science, The University of Tokyo, Bunkyo-ku, Tokyo 113-0033, Japan \\
$^{3}$Sydney Institute for Astronomy, School of Physics, The University of Sydney, NSW 2006, Australia\\
$^4$Stellar Astrophysics Centre, Department of Physics and Astronomy, Aarhus University, 8000 Aarhus C, Denmark\\
}

\begin{document}

\maketitle

\begin{abstract}
There are many Slowly Pulsating B (SPB) stars and $\gamma$~Dor stars in the {\it Kepler} Mission data set. The light curves of these pulsating stars have been classified phenomenologically into stars with symmetric light curves and with asymmetric light curves. In the same effective temperature ranges as the $\gamma$~Dor and SPB stars, there are variable stars with downward light curves that have been conjectured to be caused by spots. Among these phenomenological classes of stars, some show `frequency groups' in their amplitude spectra that have not previously been understood. While it has been recognised that nonlinear pulsation gives rise to combination frequencies in a Fourier description of the light curves of these stars, such combination frequencies have been considered to be a only a minor constituent of the amplitude spectra. In this paper we unify the Fourier description of the light curves of these groups of stars, showing that many of them can be understood in terms of only a few base frequencies, which we attribute to g~mode pulsations, and combination frequencies, where sometimes a very large number of combination frequencies dominate the amplitude spectra. The frequency groups seen in these stars are thus tremendously simplified. We show  observationally that the combination frequencies can have amplitudes greater than the base frequency amplitudes, and we show theoretically how this arises. Thus for some $\gamma$~Dor and SPB stars combination frequencies can have the highest observed amplitudes. Among the B stars are pulsating Be stars that show emission lines in their spectra from occasional ejection of material into a circumstellar disk. Our analysis gives strong support to the understanding of these pulsating Be stars as rapidly rotating SPB stars, explained entirely by g~mode pulsations. 
\end{abstract}


\section{Introduction}
\label{sec:intro}

It is well known that Fourier analysis of non-sinusoidal light curves gives rise to harmonics and combination frequencies that describe the light curve shape in terms of sinusoids. High amplitude pulsations are nonlinear, giving rise to significant amplitudes at the harmonics of the base frequencies. Multi-mode nonlinear pulsation results in interaction among the base frequencies and their harmonics that give rise to sum and difference combination frequencies of the form $n f_i \pm m f_j$. Studying the relationships among the amplitudes and phases of the Fourier components has been standard practice for RR\,Lyr stars and Cepheids since the pioneering work of \citet{simonlee81}, and the study of combination frequencies and their astrophysical implications is well established for white dwarf stars (see, e.g. \citealt{wu2001} and \citealt{montgomery2005}). Combination frequencies dominate the amplitude spectra of some $\delta$~Sct stars, for example KIC\,11754974 \citep{murphyetal2013} and KIC\,8054146 (\citealt{bregeretal2012}; \citealt{bregermontgomery2014}), where the astrophysical implications and uses of the combination frequencies are more uncertain than for white dwarf stars. 

Among B, A and F main-sequence stars there are two classes of g~mode pulsators: the $\gamma$~Dor stars with temperatures in the range of early to mid-F stars, and the Slowly Pulsating B (SPB) stars with temperatures in the range of the late B stars. The light curves of these stars have been widely discussed phenomenologically, particularly in the era of the photometric space missions MOST, CoRoT and {\it Kepler}. \citet{debosscheretal2011} performed an automated variability analysis on about 150\,000 light curves from the {\it Kepler} Quarter 1 data, finding many $\gamma$~Dor stars. \citet{balonaetal2011a} visually scanned about 10\,000 stars in the {\it Kepler} data in the temperature range of the $\gamma$~Dor stars and the coolest SPB stars and classified the light curves phenomenologically as symmetric or asymmetric, where the asymmetric light curves show larger range at maximum amplitude than at minimum amplitude. \citet{tkachenkoetal2013} determined atmospheric parameters from high resolution spectra for 69 stars in the {\it Kepler} data set that have $\gamma$~Dor g-mode pulsations. \citet{balonaetal2011b} similarly provided visual descriptions of 48 B stars in the {\it Kepler} data. Following the lead of these papers, \citet{mcnamaraetal2012} classified the light curves of 252 B stars in the {\it Kepler} data, describing many of the stars as `Fg', meaning that they show frequency groups in their amplitude spectra. Most recently \citet{bradleyetal2015} searched among 2768 {\it Kepler} stars for $\gamma$~Dor stars, $\delta$~Sct stars and so-called `hybrid' stars that show both p-mode and g-mode pulsations. They adopted the notation of \citet{balonaetal2011a} to describe the light curves as `symmetric' and `asymmetric'. 

Many of these papers used limited data sets from {\it Kepler}, unintentionally resulting in significant confusion in the description of the light curves. In addition to pulsation, stars may show light variability caused by orbital or rotational variations. Those are typically non-sinusoidal, hence also give rise to harmonics of the base frequencies. They do not, however, generate combination frequencies. It is thus possible to distinguish pulsation from rotational or orbital variability when combination frequencies are present. 

From the unprecedented time span of 4\,yr  of the full {\it Kepler} data set, it is now clear that g~modes in $\gamma$~Dor stars and SPB stars can be so closely spaced in frequency that data sets spanning less than 1\,yr may not resolve the individual pulsation frequencies. Excellent examples of this are seen in the $\gamma$~Dor -- $\delta$~Sct stars KIC~11145123 (\citealt{kurtzetal2014}; \citealt{vanreethetal2015}) and KIC~9244992 \citep{saioetal2015}, and in several other examples given by \citet{beddingetal2014} and \citet{vanreethetal2015}, where the frequency spacings of long series of consecutive radial overtone g~modes with rotational multiplets require up to half a year of data for full resolution. Thus previous descriptions of the character of $\gamma$~Dor and SPB light curves based on relatively short data sets should be viewed with caution. The visual descriptions of light curves where pulsation modes are not resolved and combination frequencies are not recognised have led to erroneous conclusions. 

The presence of combination frequencies in $\gamma$~Dor and SPB stars has been recognised by many. \citet{degrooteetal2009} developed an automated combination frequency search for CoRoT SPB stars. \citet{papics2012} discussed in general the search for combination frequencies for B, A and F main-sequence stars and problems associated with their identification, while \citet{balona2012} gave a detailed discussion in the case of $\delta$~Sct stars, particularly in comparison with white dwarf stars with the purpose of eventually using the information in the combination frequencies for astrophysical inference. 

Nevertheless, as \citet{balona2012} pointed out, the combination frequencies have usually been considered a nuisance in the search for pulsation mode frequencies for asteroseismology. He also states that `combination frequencies are of much lower amplitude than parent mode $\ldots$ [frequencies] $\ldots$'. While this may be a widely held view, it is not necessarily true: {\it Combination frequencies can have {observed} amplitudes greater than those of the base frequencies, as we show theoretically  in Section\,\ref{sec:theory}.} 

Another misconception is that a light curve that has larger variation at minimum than maximum cannot be purely pulsational. This idea has led to interpretations of some SPB and pulsating Be star light curves as being caused by spots, either completely or partially. The idea has crossed over into the visual description of the light curves of $\gamma$~Dor stars such that the papers using the terminology `asymmetric' for the non-sinusoidal pulsators only include those stars that show more variation at maximum light than minimum light, even though there are $\gamma$~Dor stars that do the opposite, as is common for SPB stars. We show examples in sections below.

We use our own description of the light curves at some expense of proliferating nomenclature. We describe stars that have non-sinusoidal light curves with larger range at maximum light than minimum light as having `upward' light curves, and those that do the opposite as having `downward' light curves. Stars previously classified as having symmetric light curves are part of a continuum between these extremes. Below we show examples of the various shapes of the light curves and their simple explanation in terms of non-sinusoidal pulsation in only a few pulsation modes with combination frequencies. We find that many $\gamma$~Dor, SPB and pulsating Be star light curves are far simpler than has previously been understood, and we make a strong case that the only physics needed to understand all of these stars is nonlinear pulsation theory. There is no need of, and no evidence for, spots. 

The reduction that we demonstrate in the apparent complexity of the amplitude spectra of the stars showing frequency groups is stunning. Instead of hundreds of frequencies being extracted for analysis, many of these stars have but a few pulsation mode frequencies with a plethora of combination frequencies, some of which can have amplitudes greater than the base frequencies. As in the cases of the p-mode pulsation in the $\delta$~Sct stars KIC\,11754974 and KIC\,8054146 mentioned above, the amplitude spectra of the g-mode pulsators on the main-sequence can be dominated by combination frequencies. These must be fully modelled to get to the pulsation mode frequencies that are the fundamental data of asteroseismology, and they have the potential to provide new astrophysical information for main-sequence stars, as they do for pulsating white dwarf stars.

In this paper we explain that what previously appeared to be complex variability with dozens or hundreds of frequencies is a result of only a few base  frequencies and their combination frequencies. This is an important observational result that greatly simplifies our understanding of the light curves of $\gamma$~Dor, SPB and pulsating Be stars. That such strong nonlinear interaction exists indicates high amplitudes for the base modes in the stellar cores. It is a goal to gain asteroseismic inference from these modes by modelling them.

\section{Theoretical corroboration} 
\label{sec:theory}

\subsection{Weak nonlinear system}

To draw some basic characteristics of nonlinear pulsation of stars, let us consider a case of weakly nonlinear pulsation, in which the  eigenfrequencies are still close to those obtained by linear calculation. We consider the unperturbed static equilibrium state of a star and superimpose on it perturbations. To make the problem simple, we assume that in the equilibrium state the star is spherically symmetric.

We define the displacement vector, $\mbox{\boldmath$\xi$}$,
\begin{eqnarray}
\lefteqn{
	\mbox{\boldmath$\xi$} (\mbox{\boldmath$r$}_0,t):= \mbox{\boldmath$r$}-\mbox{\boldmath$r$}_0,
}
\label{eq:2.1}
\end{eqnarray}
where $\mbox{\boldmath$r$}$ denotes the Lagrangian position variable of a given fluid element which is at $\mbox{\boldmath$r$}=\mbox{\boldmath$r$}_0$ in the equilibrium state.  
The equation of oscillations, which is expressed with a single variable $\mbox{\boldmath$\xi$}$, is then divided into the linear operator 
${\cal L}(\mbox{\boldmath$\xi$})$ 
and the nonlinear operators ${\cal N}^{(k)}$ $(k=2, 3, \cdots)$: 
\begin{eqnarray}
\lefteqn{
	{{\partial^2\mbox{\boldmath$\xi$}}\over{\partial t}^2} + {\cal L}(\mbox{\boldmath$\xi$})
	+{\cal N}^{(2)}(\mbox{\boldmath$\xi$}, \mbox{\boldmath$\xi$})
	+{\cal N}^{(3)}(\mbox{\boldmath$\xi$},\mbox{\boldmath$\xi$},\mbox{\boldmath$\xi$})
	+\cdots=0,
}
\label{eq:2.3}
\end{eqnarray}
where ${\cal N}^{(k)}$ denotes the operator of the $k$-th order of $\mbox{\boldmath$\xi$}$.
Retaining only the first-order terms, we obtain the linearised equation
\begin{eqnarray}
\lefteqn{
	{{\partial^2\mbox{\boldmath$\xi$}}\over{\partial t}^2} + {\cal L}\left(\mbox{\boldmath$\xi$}\right)=0.
}
\label{eq:2.4}
\end{eqnarray}
Since ${\cal L}$ does not include any operator with respect to time, 
the solution to equation (\ref{eq:2.4}), defined as $\mbox{\boldmath$\xi$}^{(1)}(\mbox{\boldmath$r$}_0,t)$, is separated into a spatial function and a temporal function. The latter is expressed by $\exp {\rm i}\omega t $, with frequency $\omega$.
Equation (\ref{eq:2.4}) turns into an eigenvalue problem with a set of suitable boundary conditions.
The eigenfunctions form an orthogonal complete set, hence the linear adiabatic oscillations of a star can be expressed as
\begin{eqnarray}
\lefteqn{
	\mbox{\boldmath$\xi$}^{(1)} (\mbox{\boldmath$r$}_0, t) = \sum_{\boldsymbol{k}} 
	a_{\boldsymbol{k}} \mbox{\boldmath$\Xi$}_{\boldsymbol{k}}^{(1)}(\mbox{\boldmath$r$}_0)\exp {\rm i}(\omega_{\boldsymbol{k}}t+\varphi_{\boldsymbol{k}}),
}
\label{eq:2.5}
\end{eqnarray}
where $\omega_{\boldsymbol k}$ and $\mbox{\boldmath$\Xi$}(\mbox{\boldmath$r$}_0)$ denotes the eigenfrequency 
and the eigenfunction of the mode index  $\boldsymbol{k}$, respectively, and $a_{\boldsymbol k}$ and $\varphi_{\boldsymbol k}$ are the amplitude and the phase of the mode at $t=0$. Here, the mode index $\boldsymbol{k}$ consists of the spherical degree $l$, the azimuthal order $m$, and the radial order $n$, and the eigenfunction $\mbox{\boldmath$\Xi$}_{\boldsymbol{k}}^{(1)}(\mbox{\boldmath$r$}_0)$ is written with the spherical coordinates $(r, \theta, \phi)$ as


\begin{eqnarray}
\lefteqn{
        \mbox{\boldmath$\Xi$}_{\boldsymbol{k}}^{(1)} (\mbox{\boldmath$r$}_0)
}
\nonumber\\
\lefteqn{
\quad
        =\Xi_{n,l}(r)Y_l^m{\bf e}_r
        + H_{n,l}(r)\left[{{\partial Y_l^m}\over{\partial\theta}} {\bf e}_{\theta}
        + {{1}\over{\sin\theta}}{{\partial Y_l^m}\over{\partial\phi}} {\bf e}_{\phi}\right],
}
\end{eqnarray}
where $Y_l^m(\theta, \phi)$ denotes the spherical harmonics with the spherical degree $l$ and the azimuthal order $m$, and $\Xi_{n,l}(r)$ and $H_{n,l}(r)$ are the radial eigenfunctions, with respect to $r$, for the displacement in the radial direction and for that in the horizontal direction, respectively, with the radial order $n$ and the spherical degree $l$.
The characteristics of the linear adiabatic pulsations of stars have already been investigated in detail, as in the textbooks  \citet{unnoetal1989} and \citet{aertsetal2010}.

Including the second-order terms of equation (\ref{eq:2.3}), we obtain the following equation for $\mbox{\boldmath$\xi$}^{(2)}(\mbox{\boldmath$r$}, t)$:
\begin{eqnarray}
\lefteqn{
        {{\partial^2\mbox{\boldmath$\xi$}^{(2)} }\over{\partial t}^2}
        + {\cal L}\left(\mbox{\boldmath$\xi$}^{(2)}\right)
        =
        -{\cal N}^{(2)}\left(\mbox{\boldmath$\xi$}^{(1)}, \mbox{\boldmath$\xi$}^{(1)}\right).
}
\label{eq:2.7}
\end{eqnarray}
Since $\mbox{\boldmath$\xi$}^{(1)}(\mbox{\boldmath$r$}_0, t)$ has already been independently solved, the above equation (\ref{eq:2.7}) is regarded as an inhomogeneous equation for $\mbox{\boldmath$\xi$}^{(2)}(\mbox{\boldmath$r$}_0, t)$ with a source term ${\cal N}^{(2)}(\mbox{\boldmath$\xi$}^{(1)}, \mbox{\boldmath$\xi$}^{(1)})$, which originates from the squared terms of the first-order free oscillations.
In other words, equation (\ref{eq:2.7}) is regarded as an equation for a forced oscillation induced by
the nonlinear term ${\cal N}^{(2)}(\mbox{\boldmath$\xi$}^{(1)}, \mbox{\boldmath$\xi$}^{(1)})$.
The particular solution to this inhomogeneous equation gives the correction to $\mbox{\boldmath$\xi$}^{(1)}(\mbox{\boldmath$r$}_0,t)$.

In a similar way, the higher-order solutions are considered as forced oscillations successively induced by the nonlinear terms of the lower-order solutions.


\subsection{Why do the combination frequencies appear?}

The operator ${\cal N}^{(2)}$ consists of cross terms of $\mbox{\boldmath$\xi$}^{(1)}(\mbox{\boldmath$r$}_0, t)$, and it is bilinear. Hence, the nonlinear term is separated into a spatial part and a temporal function. Substitution of the form of $\mbox{\boldmath$\xi$}^{(1)}$ given by equation (\ref{eq:2.5}) into equation (\ref{eq:2.7}) leads to 
\begin{eqnarray}
\lefteqn{
	{{\partial^2\mbox{\boldmath$\xi$}}\over{\partial t}^2} 
	+ {\cal L}\left(\mbox{\boldmath$\xi$}\right)
}
\nonumber \\
\lefteqn{
\quad
	=
	-\sum_{\boldsymbol{k},\boldsymbol{k'}}
	{\cal N}^{(2)} \left(a_{\boldsymbol{k}} \mbox{\boldmath$\Xi$}_{\boldsymbol{k}}^{(1)}, a_{\boldsymbol{k'}} \mbox{\boldmath$\Xi$}_{\boldsymbol{k'}}^{(1)}\right)
	{\rm e}^{{\rm i}\left\{\left(\omega_{\boldsymbol{k}}+\omega_{\boldsymbol{k'}}\right)t +\left(\varphi_{\boldsymbol{k}}+\varphi_{\boldsymbol{k'}}\right)\right\}}.
}
\label{eq:2.8}
\end{eqnarray}
The cross terms in the nonlinear operator ${\cal {N}}^{(2)}$ induce the combination frequencies. As a consequence, the particular solution to $\mbox{\boldmath$\xi$}^{(2)}(\mbox{\boldmath$r$}_0,t)$ also has combination frequencies $\omega_{\boldsymbol{k}}+\omega_{\boldsymbol{k'}}$.

It should be noted that the associated general solution of the inhomogeneous differential equation (\ref{eq:2.8}) is of the form of equation (\ref{eq:2.5}), and it is already given as a first-order solution. Hence we only have to consider the particular solution.

A special case is the cross term of $\omega_{\boldsymbol{k}}$ with itself. That induces the second harmonic $2\omega_{\boldsymbol{k}}$.
Similarly, the nonlinear operator ${\cal N}^{(3)}$ produces the third harmonic $3\omega_{\boldsymbol{k}}$ through the cross term between $\omega_{\boldsymbol{k}}$ and $2\omega_{\boldsymbol{k}}$ or the triple term of  $\omega_{\boldsymbol{k}}$. This is the process producing a non-sinusoidal light curve from a single mode.
  

\subsection{Why do some combination frequencies have amplitudes greater than their base frequencies?}

The second-order perturbation is of the order of the square of the linear perturbation. It should be noted here, however, that this statement concerns the intrinsic amplitudes. The visibility, which is highly dependent on the surface pattern of the oscillations, must be taken into account to evaluate the actual observed amplitudes.

The nonlinear operator ${\cal N}$ induces cross terms of spherical harmonics, and they are described in terms of a series of spherical harmonics with azimuthal order that is equal to the sum of the parent spherical harmonics;  
\begin{eqnarray}
\lefteqn{
	Y_l^m(\theta,\phi) Y_{l'}^{m'} (\theta, \phi) 
}
\nonumber\\
\lefteqn{
	\quad
	= \sum_{l^{''}} (-1)^{m'} c^{l''}(l,-m,l',m')Y_{l''}^{m+m'}(\theta,\phi),
}
\end{eqnarray}
where $c^{k}(l,-m,l',m')$ is defined by
\begin{eqnarray}
\lefteqn{
	c^k(l,m,l',m') := \int {\rm d}\Omega \,Y_l^m(\theta,\phi)^* Y_{l'}^{m'}(\theta,\phi) Y_k^{m-m'}(\theta,\phi),
}
\end{eqnarray}
and $l''$ is in the range of \mbox{$[\,|\,l-l'\,|, l+l'\,]$}, except the range of \mbox{$[\,0, |\,m+m'\,|\,]$}.
This means that even if the first-order perturbations associated with high degree have low {observed} amplitudes and are difficult to detect, their products may induce low degree components, e.g., $l=0$, that are detectable.

So, it is not necessarily true that combination frequencies of higher-order perturbations have smaller {observed} amplitudes than the base frequencies. {\it Some combination frequencies can have {observed} amplitudes greater than those of their base frequencies.}


\subsection{Why do some stars show downward light curves?}

\begin{figure}
\begin{center}
\includegraphics[width=\linewidth, angle=0]{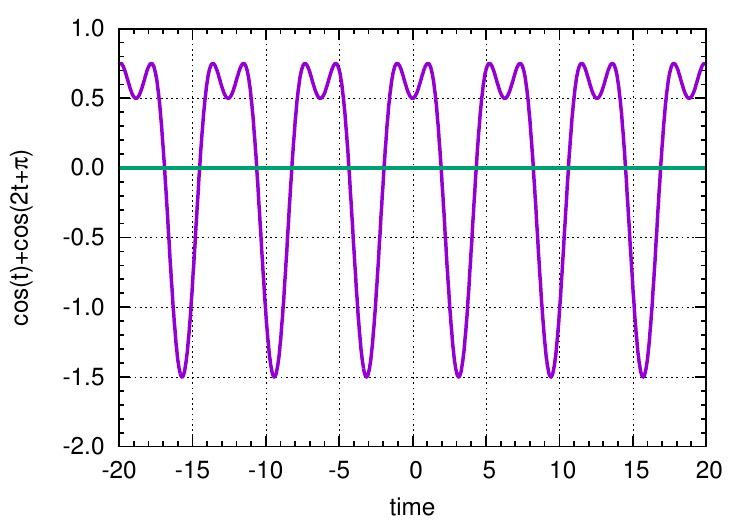}
\caption{This simulated light curve demonstrates a combination of frequencies differing by a factor of two with a phase difference of $\upi$ induces a `downward' light curve. }
\label{fig:2.1}
\end{center}
\end{figure}

Superposition of two oscillations with nearly equal frequencies leads to a beat phenomenon. With an increase in the frequency difference, the wave pattern gradually changes. When the second harmonic is imposed on the base frequency, the oscillation pattern significantly deviates from symmetry with respect to the zero level. If the phase difference of these two frequencies, $\varphi_{\boldsymbol{k}} - \varphi_{\boldsymbol{k}'}$, is nearly zero, the wave pattern shows an `upward' shape (which has previously been called `asymmetric'), whereas when the phase difference is close to $\upi$, the wave pattern shows a `downward' shape as shown in Fig.\,\ref{fig:2.1}. 

It has been widely considered that nonlinear pulsation induces only upward light curves and that pulsation cannot be the sole physical cause of downward light curves. However, as demonstrated here, pulsation induces downward light curves in some cases.  The base frequencies are driven by the $\kappa$-mechanism or by convective blocking, while the harmonic frequencies are damped by heat loss. 

Differences in thermal properties may cause phase differences between these two extreme groups to differ by $\upi$, with intermediate phases giving rise to less extreme distortion of the light curves. We thus propose that the range of pulsational light curves in B, A and F main-sequence stars, from upward through symmetric to downward shapes, is a consequence of the phases of the nonlinear harmonic and combination frequencies, and that those phases are determined by the balance between driving and damping in each individual star.
 
\section{Data and analysis methods}

\begin{table}
\centering
\caption[]{Basic data for the stars presented in this paper. $T_{\rm eff}$ and $\log g$ are from \citet{huberetal2014}. The last column gives the section of the paper where the star is discussed.}
\begin{tabular}{rrrrcc}

\toprule

\multicolumn{1}{c}{KIC} &
\multicolumn{1}{c}{$T_{\rm eff}$} & \multicolumn{1}{c}{$\log g$ } &
\multicolumn{1}{c}{Kp} & \multicolumn{1}{c}{type}  & \multicolumn{1}{c}{Section}   \\

&\multicolumn{1}{c}{K} & \multicolumn{1}{c}{cgs units} &
\multicolumn{1}{c}{mag}  &  &\\

\midrule

5450881   & $10500 \pm 250$  & $4.1 \pm 0.2$  &  12.45  & SPB & \ref{sec:545} \\
7468196   & $6850 \pm 200$     & $4.1 \pm 0.2$  & 13.77 &  $\gamma$~Dor &  \ref{sec:746_1079}\\
8113425   & $6900 \pm 200$     & $4.3 \pm 0.1$ & 13.86 &  $\gamma$~Dor & \ref{sec:8113425} \\
10118750 & $11400 \pm 400$ & $3.7 \pm 0.3$ &  13.90 &  SPB  & \ref{sec:1011} \\
10799291 & $11100 \pm 400$ & $4.4 \pm 0.1$ & 14.98 &  SPB  & \ref{sec:746_1079}\\
11971405  & $11600 \pm 400$ & $3.7 \pm 0.4$  & 9.32  & SPB/Be & \ref{sec:1197} \\

\bottomrule

\end{tabular}
\label{tab:data}
\end{table}

We have visually examined {\it Kepler} light curves and amplitude spectra for thousands of B, A and F main-sequence stars and selected examples to illustrate our results. The data used for the analysis in this paper are the {\it Kepler} quarters 0 to 17 (Q0 -- Q17) long cadence (LC) data. The {\it Kepler} `quarters' were of variable time span that depended on operational constraints. Most quarters are close to one fourth of a {\it Kepler} orbital period of 372.455~d, which was the time scale on which the satellite was rolled to keep the solar panels fully illuminated. Q0, Q1 and Q17 were short `quarters'. We do not use any stars in this paper that fell on the failed module 3, so there are no large gaps in our data sets. 

We used the multi-scale, maximum a posteriori (msMAP) pipeline data converted to magnitudes; information on the reduction pipeline can be found in the data release notes\footnote{https://archive.stsci.edu/kepler/data\_release.html} 21. To optimise the search for exoplanet transit signals, the msMAP data pipeline removes some astrophysical signals with frequencies less than 0.1\,d$^{-1}$ (or periods greater than 10\,d). Some of the combination frequencies that we find in this paper are at frequencies less than this 0.1\,d$^{-1}$ limit, and we show that these frequencies are unperturbed by the pipeline reductions. A useful general conclusion is that while the pipeline may reduce astrophysical amplitude, it does not perturb the frequencies. 

For all stars we visually inspected the light curve and removed a few obvious outliers. The time span for Q0--17 data is 1470.5\,d, and for Q1--Q17 data is 1459.5~d. Typically, about 64\,840 data points comprised each data set. Table\,\ref{tab:data} gives basic data for the stars analysed in this paper. 

\subsection{Frequency analysis}

We first produced a catalogue of all stars in the {\it Kepler} data with effective temperatures in the {\it Kepler} Input Catalogue (KIC) above 6400\,K. Our catalogue comprises light curves and amplitude spectra for each quarter of {\it Kepler} data for each star. We have visually studied each of these plots. For stars that we studied in more detail, we first examined the entire data sets, Q0--Q17, using the interactive light curve and amplitude spectrum tools in the programme {\small PERIOD04} \citep{lenz&breger2004}. We then used a Discrete Fourier Transform \citep{kurtz85} and our own least-squares and nonlinear least-squares fitting programmes to find the frequencies, amplitudes and phases to describe the light curves. After normalising the entire data set to zero in the mean, we fitted a cosine function, $\Delta m = A \cos (2 \pi f (t - t_0) + \phi)$, to the data in magnitudes, thus defining our convention for the phases in this paper. Our routines and {\small PERIOD04} are in agreement. 

Our procedure was to identify `base frequencies' in the amplitude spectrum from which to generate the combination frequencies. Those were then optimised by fitting them simultaneously by non-linear least-squares to the data. For reasons of space and presentation we do not tabulate the individual frequency uncertainties. Those depend on the signal-to-noise ratio \citep{montgomery-odonoghue99} and are in general of the order of $10^{-7} - 10^{-6}$\,d$^{-1}$. This is significantly less than the resolution of the Fourier peaks of $R \sim 1/  \Delta T = 0.0007$\,d$^{-1}$, where $T = 1460$\,d is the time span of the data, but the frequencies are much better determined than the resolution of the data set, so long as there are no unresolved frequencies. We discuss both cases in the sections below.

Importantly, following the determination of the base frequencies, we did not extract large numbers of peaks in the amplitude spectra, then test whether peaks were combination of the base frequencies. It is so clear from first inspection that the amplitude spectra of many $\gamma$~Dor and SPB stars are dominated by combination frequencies that we selected just a few base frequencies and then {\it calculated} the frequencies of the combination terms and fitted that calculated set of frequencies by least-squares to the data. The success of that procedure in accounting for most of the variance in the data sets is apparent in the sections below and justifies the method.

Of course, because of the nature of combination frequencies, other sets of base frequencies chosen from the combination frequencies can produce the same set of frequencies to be fitted to the data. Hence the identification of base frequencies with an astrophysical cause is open to interpretation. We hypothesise in this paper that all of the chosen base frequencies arise from g~mode pulsations. Other interpretations, should they arise, do not change the mathematical fit of the chosen cosinusoids to the data.

\section{The $\mbox{\boldmath$\gamma$}$~Dor star KIC\,8113425}
\label{sec:8113425}

\begin{figure}
\centering	
\includegraphics[width=0.99\linewidth,angle=0]{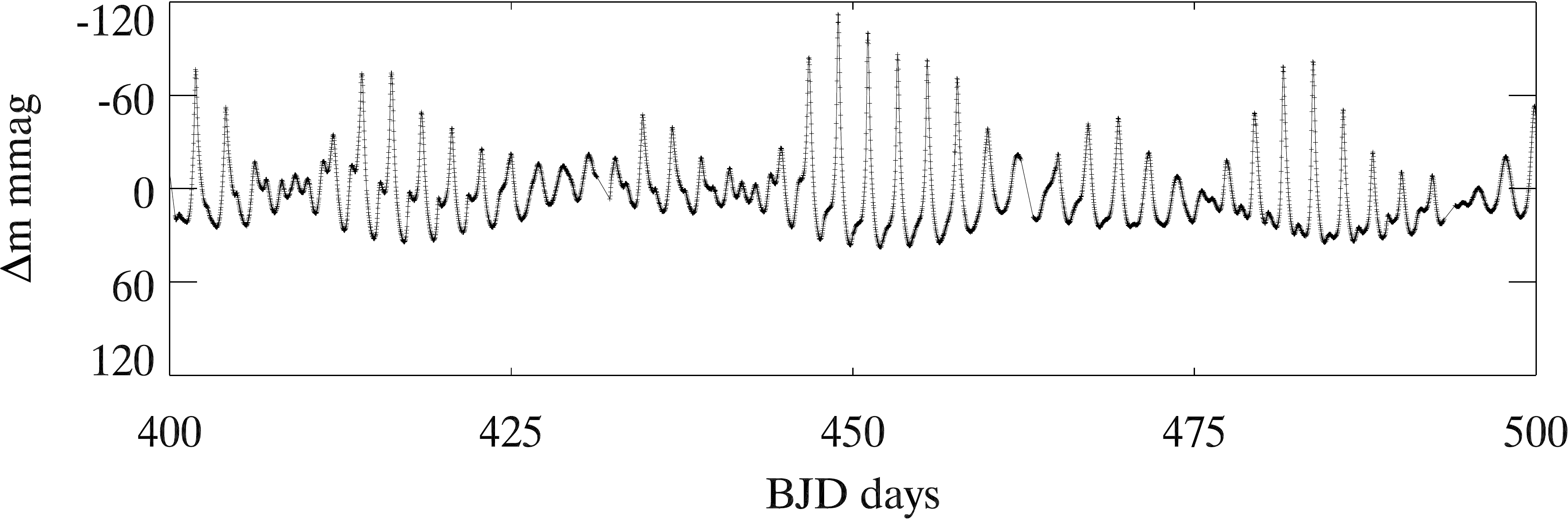}	
\caption{A section of the light curve of KIC\,8113425 spanning 100\,d showing the strongly nonlinear `upward' light variations. The time is relative to ${\rm BJD}\,245\,5000$. Almost all of the variation in the light curve is explained by the nonlinear interaction of only four g~modes.}
\label{fig:811_lc}
\end{figure}

Frequency groups were noted in some {\it Kepler} stars by \citet{mcnamaraetal2012} following the terminology of \citet{balonaetal2011a}, which we adopt. Our own examination of the light curves and amplitude spectra of thousands of {\it Kepler} B, A and F stars shows that many {\it Kepler} $\gamma$~Dor and SPB stars have such frequency groups with a wide variety of characteristics, as has been noted by others. For example, some illustration of this can also be seen in Figure~2 of \citet{tkachenkoetal2013} who studied 69 $\gamma$~Dor stars in the {\it Kepler} data set. 

These frequency groups may arise from different causes, but many of the stars showing them have only a few nonlinear pulsation modes with amplitude spectra that are dominated by combination frequency peaks. We illustrate this with an extreme case: KIC\,8113425 is a $\gamma$~Dor star with an upward light curve, which is shown in Fig.\,\ref{fig:811_lc} with a 100-d section of the light curve. The strongly nonlinear pulsations can be seen clearly. This star previously was listed as an `asymmetric' light curve star that shows `evidence of migrating starspots' according to \citet{balonaetal2011a}. We argue here that the star has no spots; that suggestion came from a simple visual inspection of a short section of the light curve with no frequency analysis. 

\begin{figure}
\centering	
\includegraphics[width=0.99\linewidth,angle=0]{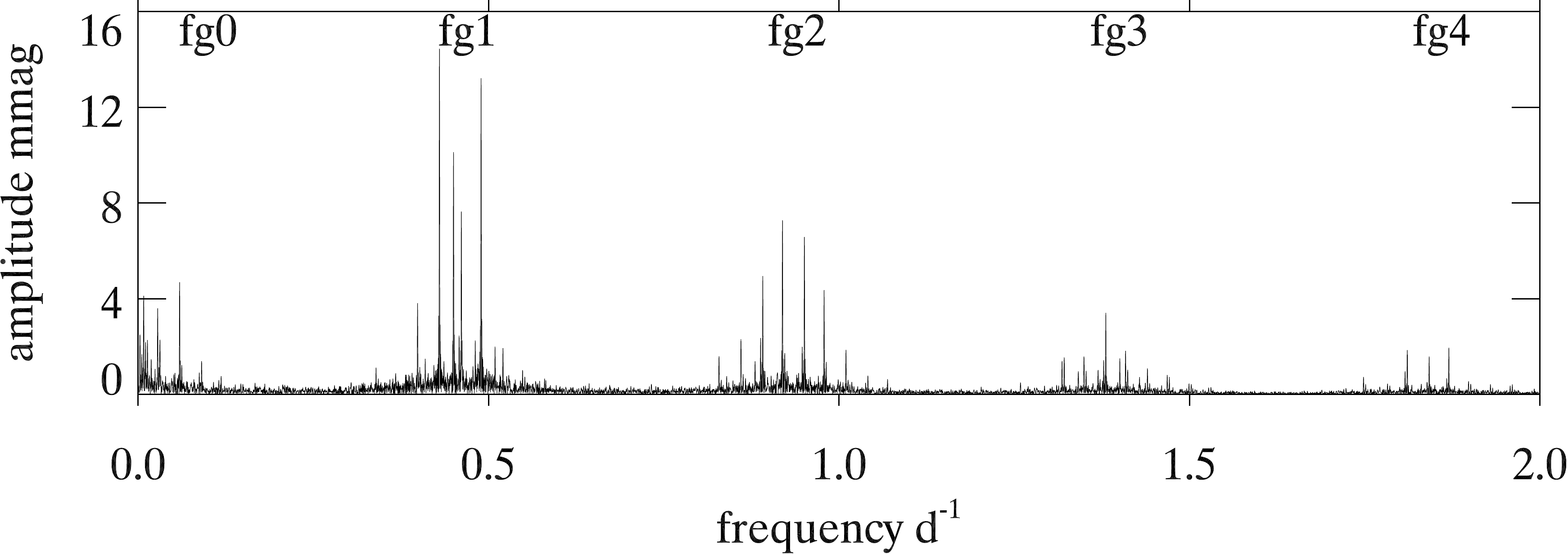}	
\includegraphics[width=0.99\linewidth,angle=0]{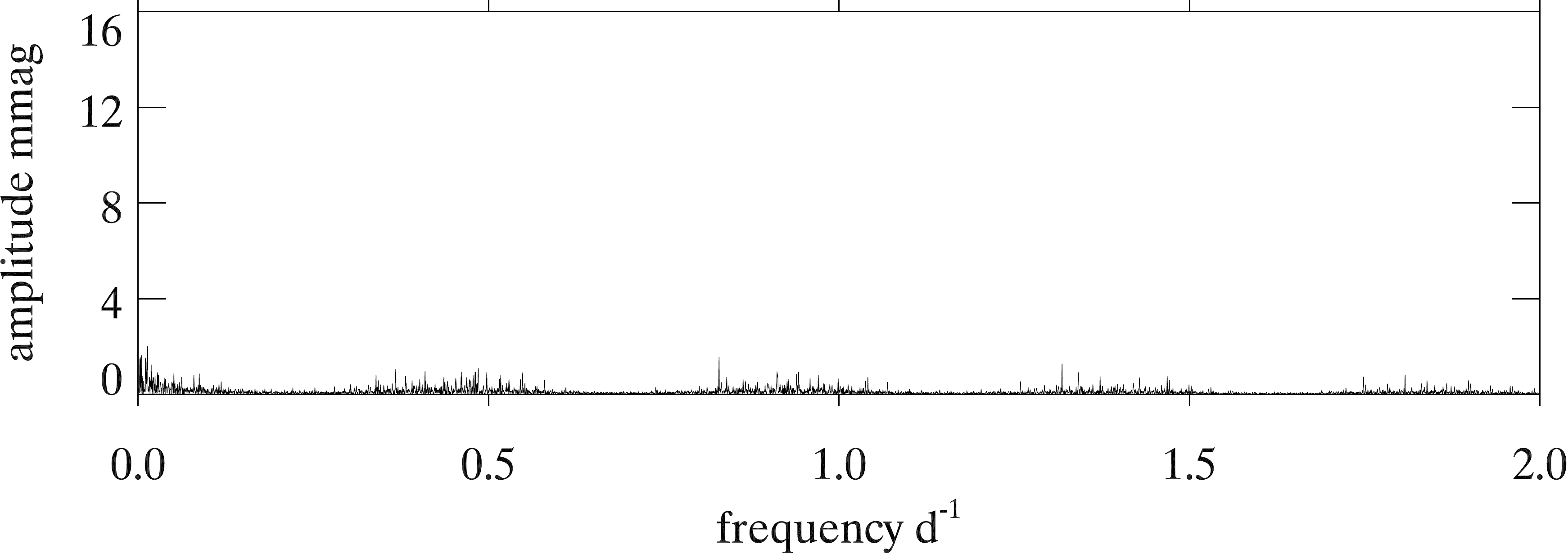}	
\includegraphics[width=0.99\linewidth,angle=0]{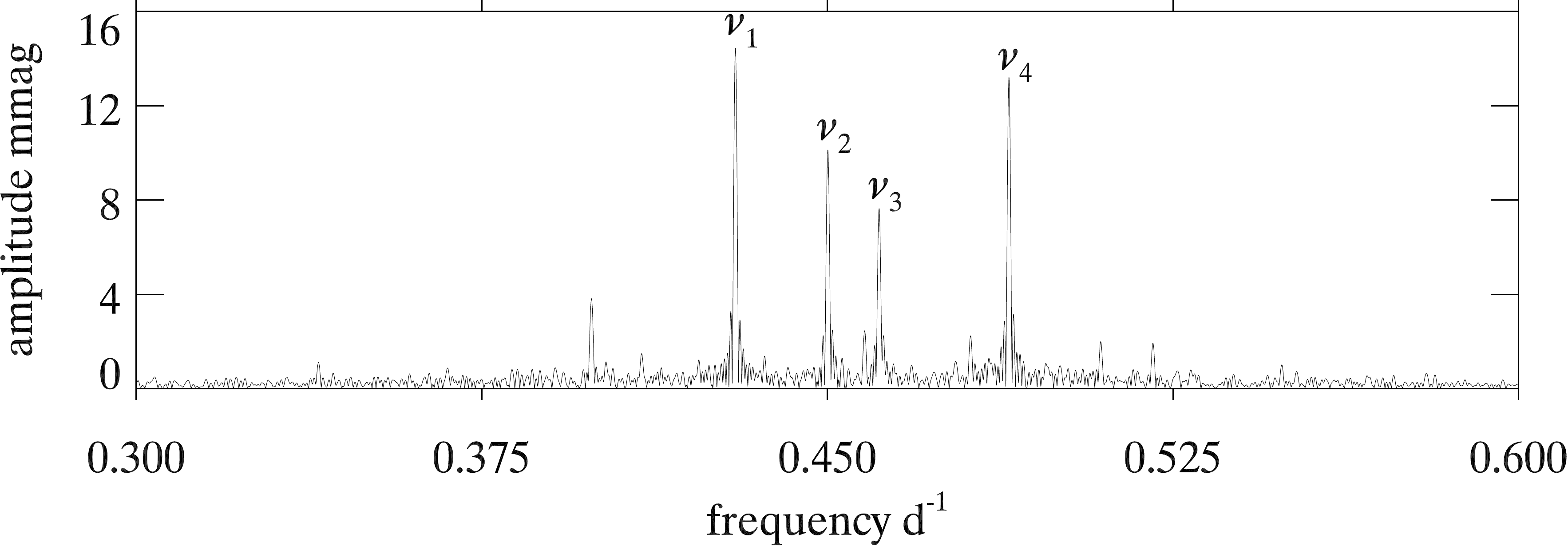}	
\caption{Top: An amplitude spectrum of the {\it Kepler} Q1-17 data for KIC\,8113425 out to 2\,d$^{-1}$. There are no p~mode pulsations at higher frequencies up to the Nyquist frequency. The frequency groups are labelled up to fg4, but can be seen to extend further; at least 11 groups can be seen in the amplitude spectrum of this star.  The middle panel shows on the same scale an amplitude spectrum of the residuals after pre-whitening the four pulsations frequencies and 39 combination frequencies with terms up to order $2\nu$ given in Table\,\ref{tab:811}, and with amplitudes greater than 1\,mmag; the inclusion of combination frequencies with amplitudes less than 1\,mmag removes more peaks.  Higher order combination frequencies explain the higher frequency groups. The reduction in variance is stunning. The four base frequencies from frequency group 1 (fg1) are shown and labelled in the bottom panel. We propose that those represent g~mode pulsations.}
\label{fig:8113425ft_all}
\end{figure}

KIC\,8113425 was observed by {\it Kepler} in LC in all quarters from Q1 to Q17. No other published observations or data are available. Fig.\,\ref{fig:8113425ft_all} shows an amplitude spectrum of the Q1--17 {\it Kepler} data out to a frequency of 2\,d$^{-1}$ where it can be seen that there are many frequency groups, which we label fg0, fg1, etc. In a close examination of the amplitude spectrum, frequency groups up to fg11 can be seen, showing the presence of very high order combination frequencies. 

We chose KIC\,8113425 to discuss the understanding of the frequency groups because it shows so many groups which are well-separated in frequency. It appears to be complex and to be an extreme example of the frequency group phenomenon, yet the apparent complexity lies almost entirely in the combination frequencies -- we have fitted most of the variance with only four base frequencies.  Table\,\ref{tab:811} lists those four base frequencies and 39 combination frequencies with terms up to order $2\nu$ that were fitted to produce the amplitude spectrum of the residuals shown in the middle panel of Fig\,\ref{fig:8113425ft_all}.

We limit our discussion here to combination frequencies with terms to order $2\nu$ and amplitudes greater than 1\,mmag so that we have a manageable number of frequencies from which to discuss the details. A fit of combination frequencies with terms up to order $5\nu$ with no lower limit on amplitude yields over 500 combination frequencies and reduces the variance in the amplitude spectrum of the residuals further.  Obviously, to explain the highest frequency groups would need even higher order combination frequencies. As \citet{papics2012} warned, with higher order fits of combination frequencies, the number of fitted peaks can become large with respect to the number of independent Fourier peaks in the amplitude spectrum. Thus in future detailed studies of these stars with frequency groups, care will need to be taken with high-order fits to avoid combination frequencies that occur only by coincidence. With the large amplitudes and low number of combination frequencies we use here, such chance coincidences are not a problem.

\begin{table}
\centering
\caption[]{A least-squares fit of the four pulsation mode frequencies of KIC\,8113425 and their combination frequencies with terms up to order $2\nu$ having amplitudes greater than 1\,mmag. There are 43 identified frequencies, including the four base frequencies.  The zero point of the time scale is ${\rm BJD} \,245\,5694.25$.}
\begin{tabular}{clrr}
\toprule
\multicolumn{1}{c}{labels} &
\multicolumn{1}{c}{frequency} & \multicolumn{1}{c}{amplitude} &
\multicolumn{1}{c}{phase} \\
&\multicolumn{1}{c}{d$^{-1}$} & \multicolumn{1}{c}{mmag} &
\multicolumn{1}{c}{radians}  \\
& & \multicolumn{1}{c}{$\pm 0.03$} &\\

\midrule
\multicolumn{4}{c}{fg0}\\
\midrule

$ -\nu_1 + 2\nu_3 -\nu_4    $ & $  0.003055   $ & $  1.87      $ & $  -1.612   \pm  0.014$  \\
$   \nu_1 -\nu_2 -\nu_3 + \nu_4    $ & $  0.008108   $ & $  4.18      $ & $  -1.757   \pm  0.007$  \\
$   -\nu_2 + \nu_3  $ & $  0.011163   $ & $  1.20      $ & $  1.088   \pm  0.025$  \\
$    -\nu_3 + \nu_4    $ & $  0.028150   $ & $  3.71      $ & $  -1.073   \pm  0.008$  \\
$  -\nu_1 + \nu_3  $ & $  0.031206   $ & $  2.42      $ & $  1.997   \pm  0.012$  \\
$  -\nu_1 + \nu_4    $ & $  0.059356   $ & $  4.32      $ & $  0.874   \pm  0.007$  \\
$  -2\nu_1 + \nu_3 + \nu_4    $ & $  0.090562   $ & $  1.37      $ & $  -0.033   \pm  0.021$  \\

\midrule
\multicolumn{4}{c}{fg1}\\
\midrule

$   2\nu_1 -\nu_3  $ & $  0.398853   $ & $  3.66      $ & $  1.299   \pm  0.008$  \\
$   2\nu_1 -\nu_2  $ & $  0.410016   $ & $  1.02      $ & $  2.107 \pm  0.030$  \\
$    \nu_2 + \nu_3 -\nu_4    $ & $  0.421951   $ & $  1.04      $ & $  2.913   \pm  0.029$  \\
$   \nu_1  $ & $  0.430058   $ & $  14.55      $ & $  0.742  \pm 0.002 $  \\
$    \nu_2  $ & $  0.450101   $ & $  10.09      $ & $  -0.449   \pm  0.003$  \\
$   \nu_1 -\nu_3 + \nu_4    $ & $  0.458209   $ & $  1.95      $ & $  -2.793   \pm  0.015$  \\
$    \nu_3  $ & $  0.461264   $ & $  7.75      $ & $  -1.995   \pm  0.004$  \\
$  -\nu_1 + \nu_2 + \nu_3  $ & $  0.481307   $ & $  1.73      $ & $  1.891   \pm  0.017$  \\
$    \nu_4    $ & $  0.489414   $ & $  13.14      $ & $  -3.136  \pm 0.002 $  \\
$   -\nu_2 + \nu_3 + \nu_4    $ & $  0.500577   $ & $  1.57      $ & $  -0.331   \pm  0.019$  \\
$  -\nu_1 + \nu_2 + \nu_4    $ & $  0.509457   $ & $  1.47      $ & $  -2.478   \pm  0.021$  \\
$  -\nu_1 + \nu_3 + \nu_4    $ & $  0.520620   $ & $  1.51      $ & $  2.294   \pm  0.020$  \\

\midrule
\multicolumn{4}{c}{fg2}\\
\midrule

$   2\nu_1  $ & $  0.860117   $ & $  2.42      $ & $  -1.921   \pm  0.012$  \\
$   \nu_1 + \nu_2  $ & $  0.880159   $ & $  1.21      $ & $  -2.729   \pm  0.025$  \\
$   2\nu_1 -\nu_3 + \nu_4    $ & $  0.888267   $ & $  1.91      $ & $  1.134   \pm  0.016$  \\
$   \nu_1 + \nu_3  $ & $  0.891322   $ & $  4.79      $ & $  1.129   \pm  0.006$  \\
$   \nu_1 + \nu_4    $ & $  0.919473   $ & $  7.39      $ & $  0.647 \pm 0.004   $  \\
$    2\nu_3  $ & $  0.922528   $ & $  2.16      $ & $  -2.381   \pm  0.014$  \\
$   \nu_1 -\nu_3 + 2\nu_4    $ & $  0.947623   $ & $  1.29      $ & $  -2.936   \pm  0.023$  \\
$    \nu_3 + \nu_4    $ & $  0.950678   $ & $  6.48      $ & $  -2.749   \pm  0.005$  \\
$    2\nu_4    $ & $  0.978828   $ & $  4.10      $ & $  -3.036   \pm  0.007$  \\
$  -\nu_1 + 2\nu_3 + \nu_4    $ & $  0.981884   $ & $  1.01      $ & $  2.974   \pm  0.029$  \\
$  -\nu_1 + \nu_3 + 2\nu_4    $ & $  1.010034   $ & $  1.63      $ & $  2.264   \pm  0.018$  \\

\midrule
\multicolumn{4}{c}{fg3}\\
\midrule

$   2\nu_1 + \nu_3  $ & $  1.321381   $ & $  1.50      $ & $  -1.083   \pm  0.020$  \\
$   2\nu_1 + \nu_4    $ & $  1.349531   $ & $  1.64      $ & $  -1.520   \pm  0.018$  \\
$   \nu_1 + 2\nu_3  $ & $  1.352586   $ & $  1.02      $ & $  2.184   \pm  0.029$  \\
$   \nu_1 + \nu_2 + \nu_4    $ & $  1.369574   $ & $  1.02      $ & $  -2.624   \pm  0.029$  \\
$   2\nu_1 -\nu_3 + 2\nu_4    $ & $  1.377681   $ & $  1.15      $ & $  1.409   \pm  0.026$  \\
$   \nu_1 + \nu_3 + \nu_4    $ & $  1.380736   $ & $  3.33      $ & $  1.522   \pm  0.009$  \\
$    \nu_2 + \nu_3 + \nu_4    $ & $  1.400779   $ & $  1.45      $ & $  0.119   \pm  0.021$  \\
$   \nu_1 + 2\nu_4    $ & $  1.408887   $ & $  1.88      $ & $  0.814   \pm  0.016$  \\
$    2\nu_3 + \nu_4    $ & $  1.411942   $ & $  1.04      $ & $  -1.284   \pm  0.029$  \\
$    \nu_3 + 2\nu_4    $ & $  1.440092   $ & $  1.07      $ & $  -1.964   \pm  0.028$  \\

\midrule
\multicolumn{4}{c}{fg4}\\
\midrule

$   2\nu_1 + \nu_3 + \nu_4    $ & $  1.810795   $ & $  1.87      $ & $  -0.920   \pm  0.016$  \\
$   \nu_1 + 2\nu_3 + \nu_4    $ & $  1.842000   $ & $  1.61      $ & $  2.097   \pm  0.018$  \\
$   \nu_1 + \nu_3 + 2\nu_4    $ & $  1.870151   $ & $  1.94      $ & $  1.606   \pm  0.015$  \\

\bottomrule

\end{tabular}
\label{tab:811}
\end{table}

\subsection{The phases}

The asymmetry of the light curve is described through the combination frequencies, but the contribution of each frequency to that description differs. Three factors are important in determining the asymmetry described by each combination. We explain them here for KIC\,8113425, where the upward light curve shows high maxima and shallow minima, but the description could also be applied to a downward light curve with low maxima and deep minima, or any intermediate shape light curve. The discussion can thus be generalised from this specific example.

\begin{figure}
\centering	
\includegraphics[width=0.95\linewidth]{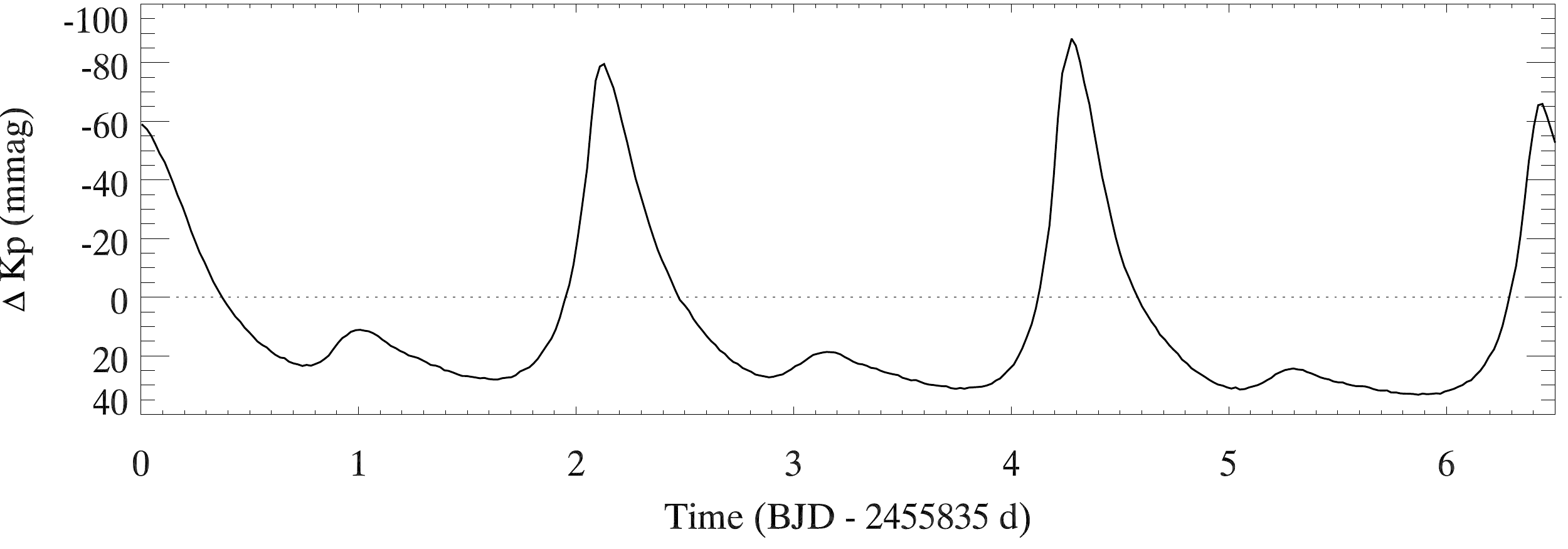}
\includegraphics[width=0.95\linewidth]{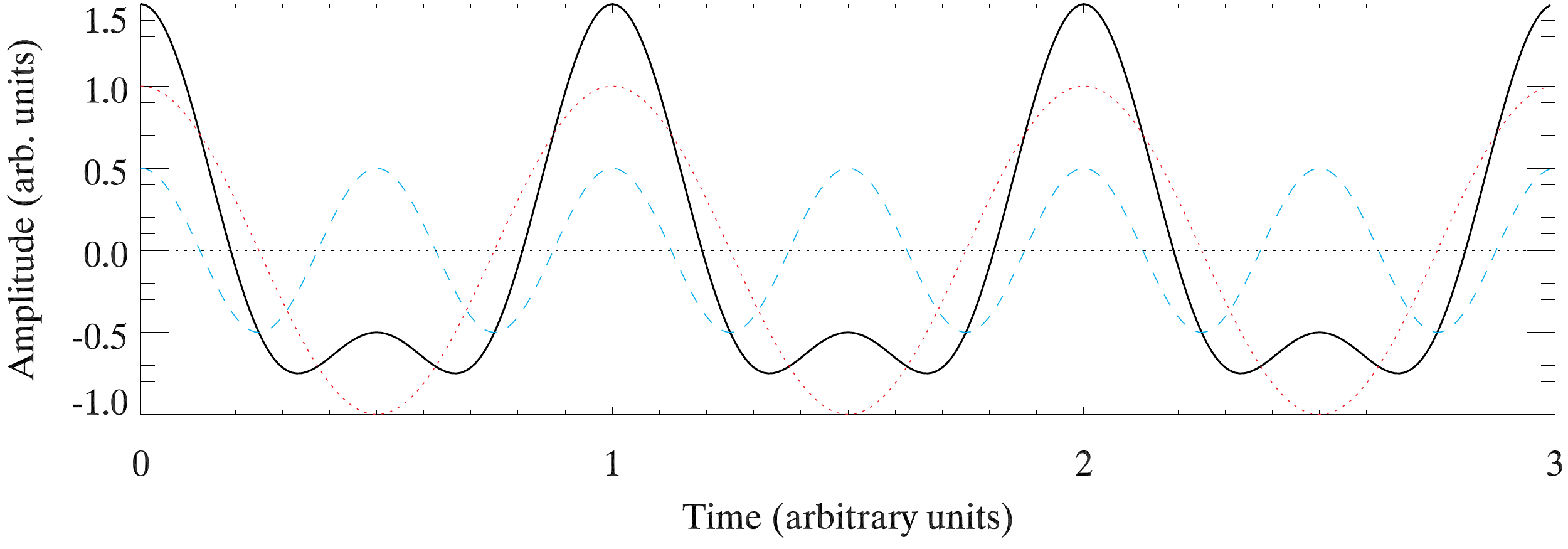}	
\caption{Top: \textit{Kepler} observations of KIC\,8113425. Bottom: an artificial light curve (black) constructed with a single frequency [red $ =\cos(2 \pi x)$] and its harmonic [blue $=0.5\cos(2 \times 2 \pi x)$]. The relative phase of the harmonic is zero, so it reinforces the maxima and suppresses the minima, leading to an `upward' light curve. Similarities with the observations in the top panel are already evident after including only one combination term.}
\label{fig:phase_factor}
\end{figure}

The first important factor is the combination frequency. In order to describe sharp and high maxima, the combination frequency must be higher than the base frequency. The second factor is the phase. To describe high maxima in the light curve, the combinations need a relative phase close to zero. The relative phase\footnote{We calculated the relative phases with a cosine function applied to the luminosity variations so that upward light curves have combination frequency phases in the positive x direction in the phasor plots, whereas, for purposes of pre-whitening in the frequency analysis where logarithms are preferred, the phases in the tables are for magnitude variations. There is thus a $\upi$\,rad shift in each phase in the tables that was used to calculate the phases in the phasor plots. }, $\phi_{\rm r}$, of a combination frequency is defined as
\begin{equation}
\phi_{\rm r} = \phi_{\rm obs} - \phi_{\rm calc} = \phi_{\rm obs} - (n\phi_i + m\phi_j)
\end{equation}
where $\phi_{\rm obs}$ is the observed phase of the combination frequency, and $\phi_{\rm calc}$ is a phase calculated from the base frequencies, which in this case is for the combination $\nu = n\nu_i + m\nu_j$. A relative phase of zero means the maximum of the combination frequency coincides with the maxima of the base frequencies, and this maximum is then reinforced. An example is shown in Fig.\,\ref{fig:phase_factor}. Finally, the third factor is amplitude. Combinations with low amplitudes have little relevance to the shape of the light curve.

\begin{figure}
\centering	
\includegraphics[width=0.9\linewidth,angle=0]{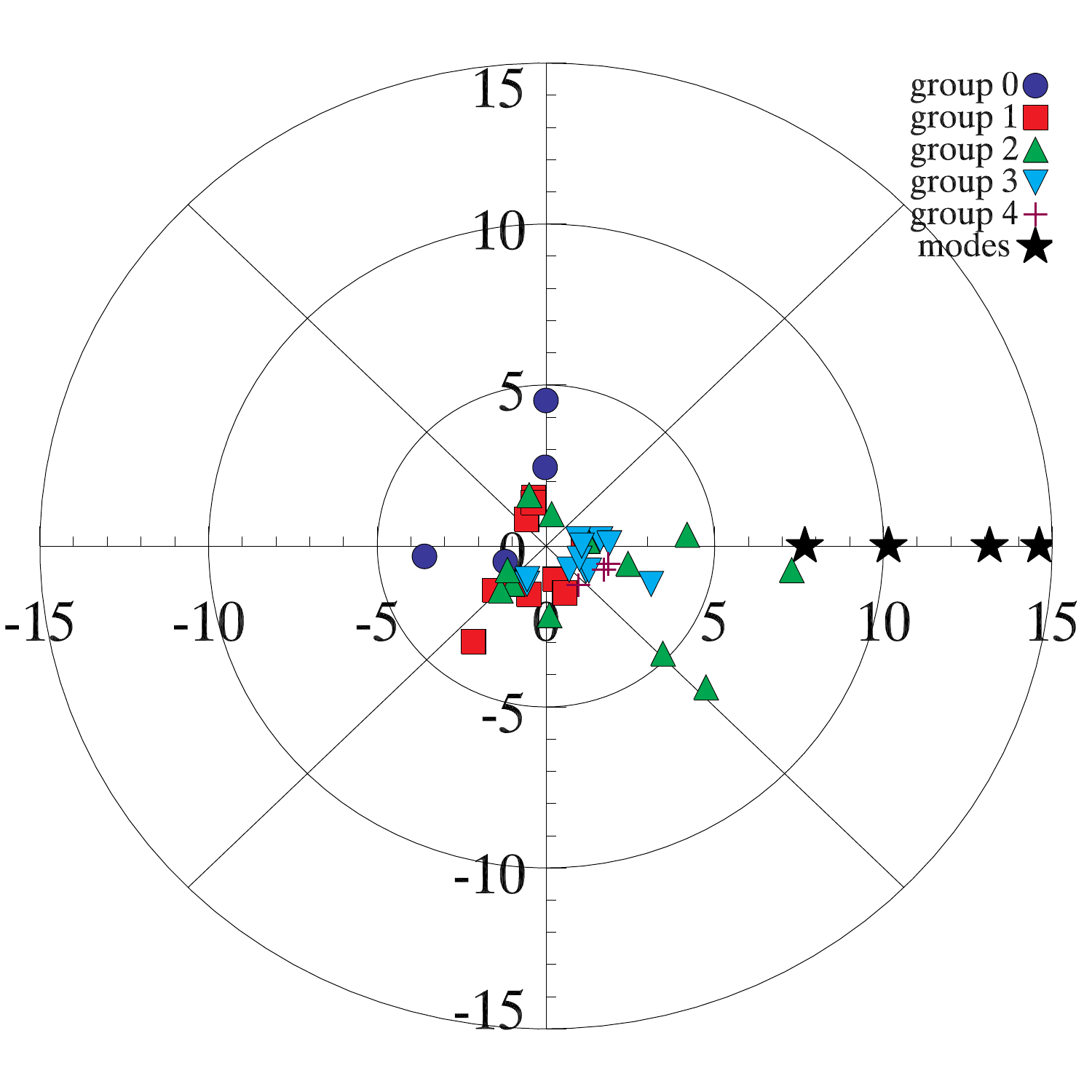}	
\caption{A phasor plot for KIC\,8113425 showing the relationship of the phases of the combination frequencies to those of the base frequencies (black stars), which by definition have relative phase zero (i.e., they lie along the positive Cartesian x-axis, which represents $\phi_r = 0$). Variables shown are amplitude and relative phase of the combination frequencies, using cosines to describe the luminosity variation in units of parts-per-thousand. Hence, points that are right of centre correspond to combination frequencies describing `upward' asymmetry. The groups described with different colours and symbols correspond to the frequency groups of Table\:\ref{tab:811}.}
\label{fig:phasor}
\end{figure}

The contribution each combination frequency makes to the description of the asymmetry can be shown in a {\it phasor} (a contraction of `phase vector') diagram, Fig.\,\ref{fig:phasor}. Here, the amplitude and relative phase of each combination are shown on a polar plot. The conventions chosen dictate the orientation of the phasor diagram; here we have chosen cosines to fit the luminosity variation, so that points that lie to the right of centre correspond to combination frequencies describing `upward' asymmetry. Similarly, points to the left of centre belong to combinations describing `downward' asymmetry. Since KIC\,8113425 has a strong upward asymmetry, we see points of high amplitude in the right part of the diagram.

\clearpage
\clearpage

\section{A $\mbox{\boldmath$\gamma$}$~Dor star and an SPB star: upward and downward light curves compared}
\label{sec:746_1079}

\begin{figure}
\centering	
\includegraphics[width=0.99\linewidth,angle=0]{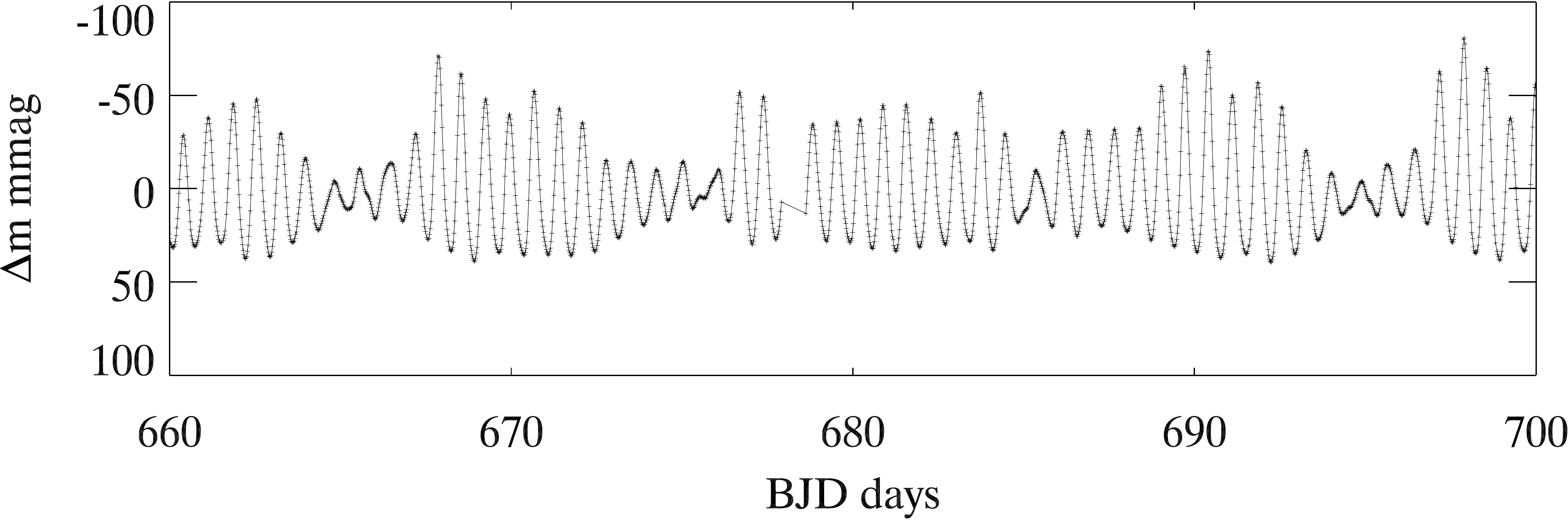}	
\includegraphics[width=0.99\linewidth,angle=0]{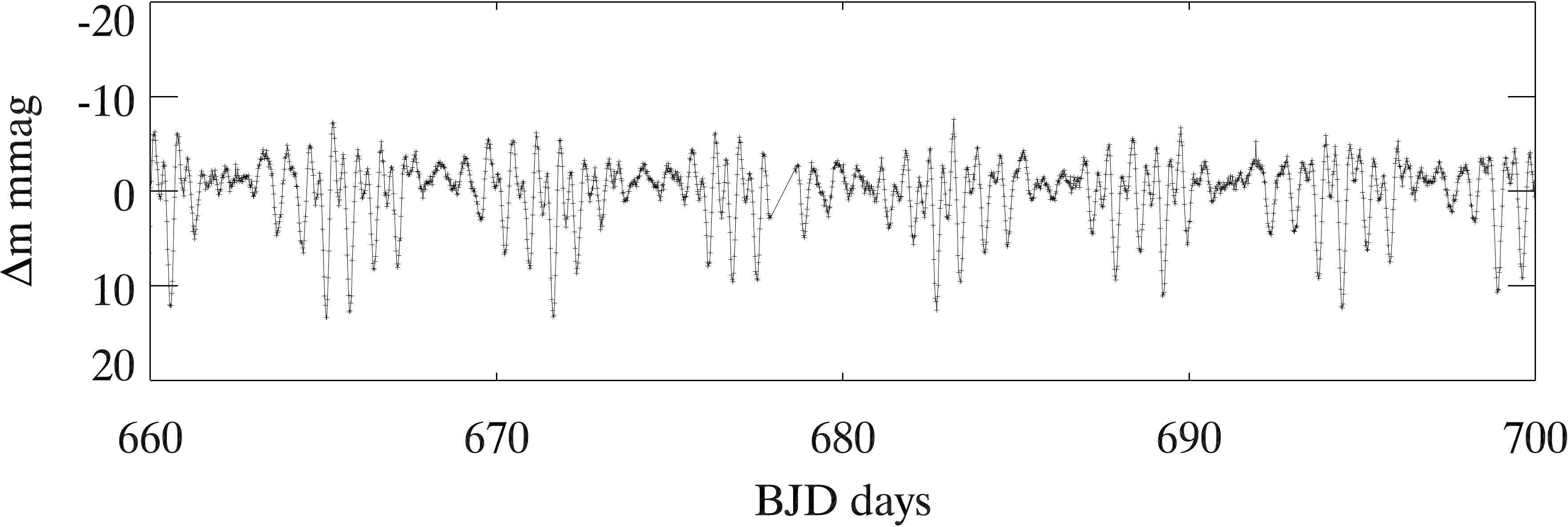}	
\caption{Top panel: A section of the light curve of the $\gamma$~Dor star KIC\,7468196 spanning 40~d showing the upward light variations. Bottom panel: A section of the light curve of the SPB star KIC\,10799291 spanning the same 40~d showing the downward light variations. The time is relative to ${\rm BJD} \,245\,5000$. Note that while the light curves have a different appearance, they are both explained by a few nonlinear g-mode pulsations with combination frequencies.}
\label{fig:1079_lc}
\end{figure}

\begin{figure*}
\centering	
\includegraphics[width=0.49\linewidth,angle=0]{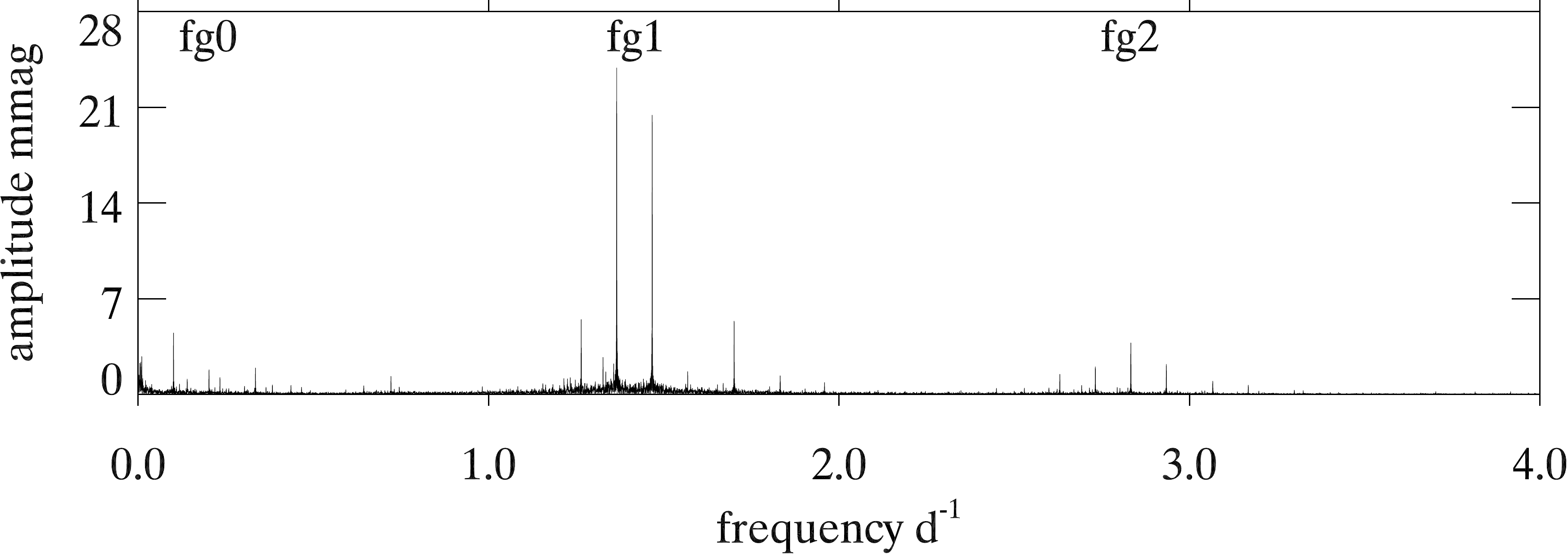}	
\includegraphics[width=0.49\linewidth,angle=0]{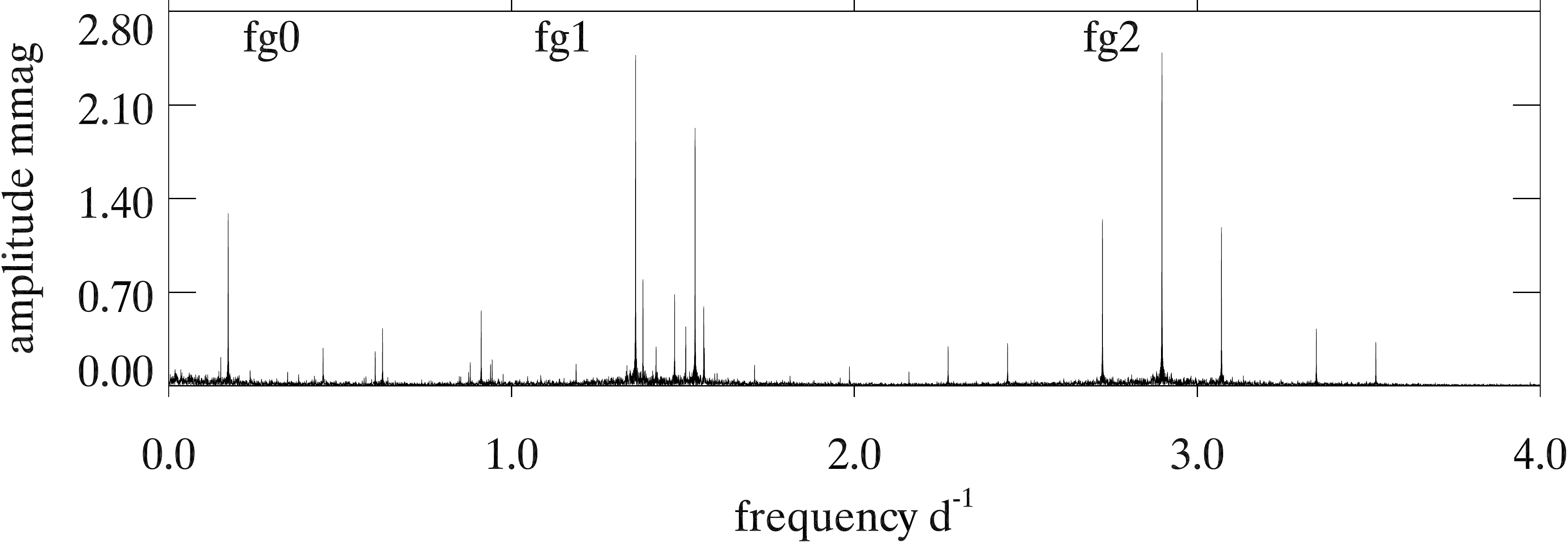}	
\includegraphics[width=0.49\linewidth,angle=0]{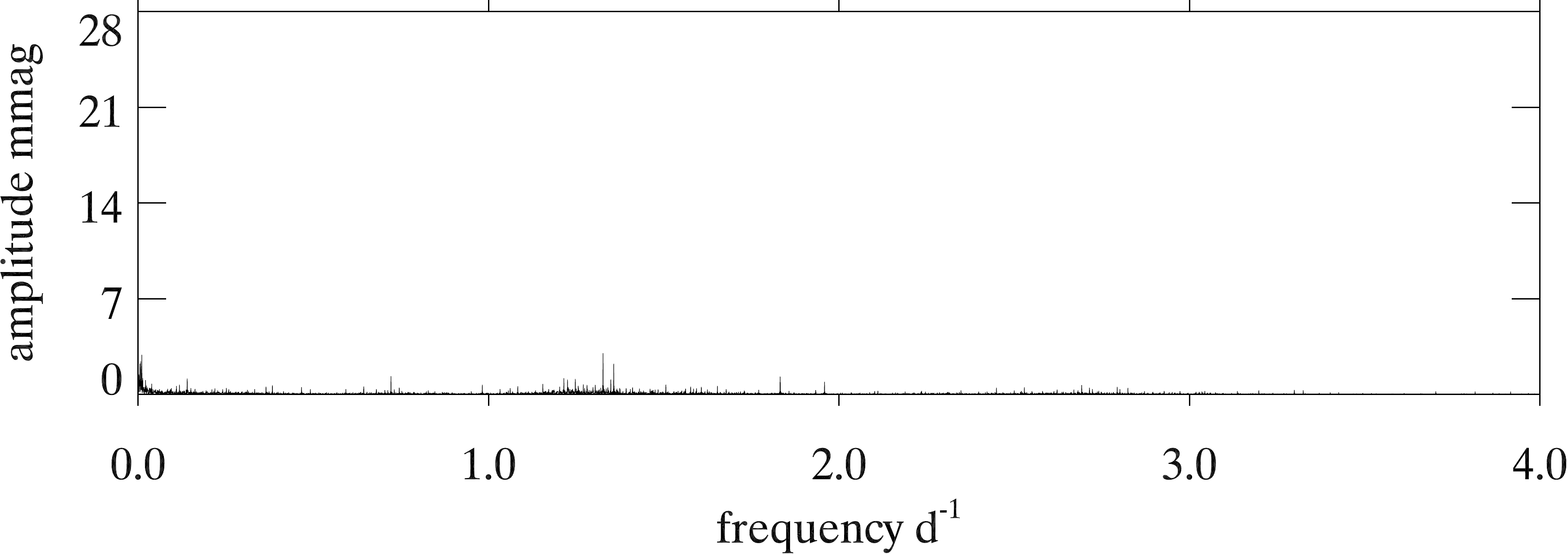}	
\includegraphics[width=0.49\linewidth,angle=0]{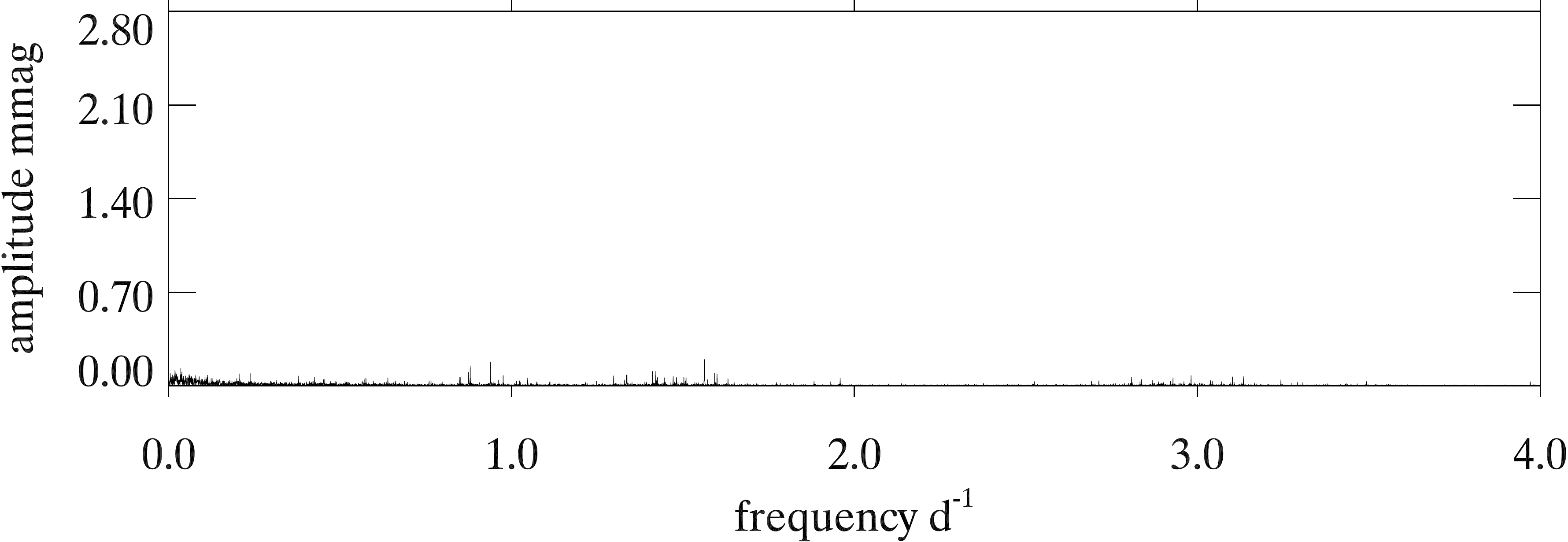}	
\includegraphics[width=0.49\linewidth,angle=0]{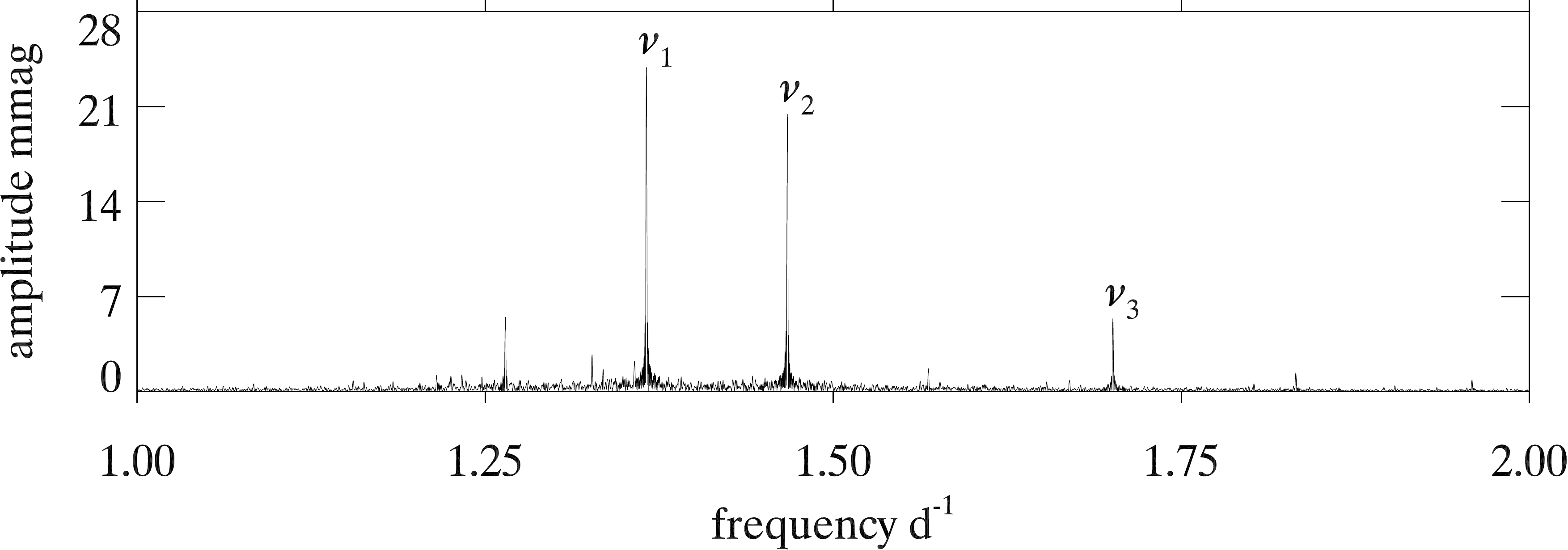}	
\includegraphics[width=0.49\linewidth,angle=0]{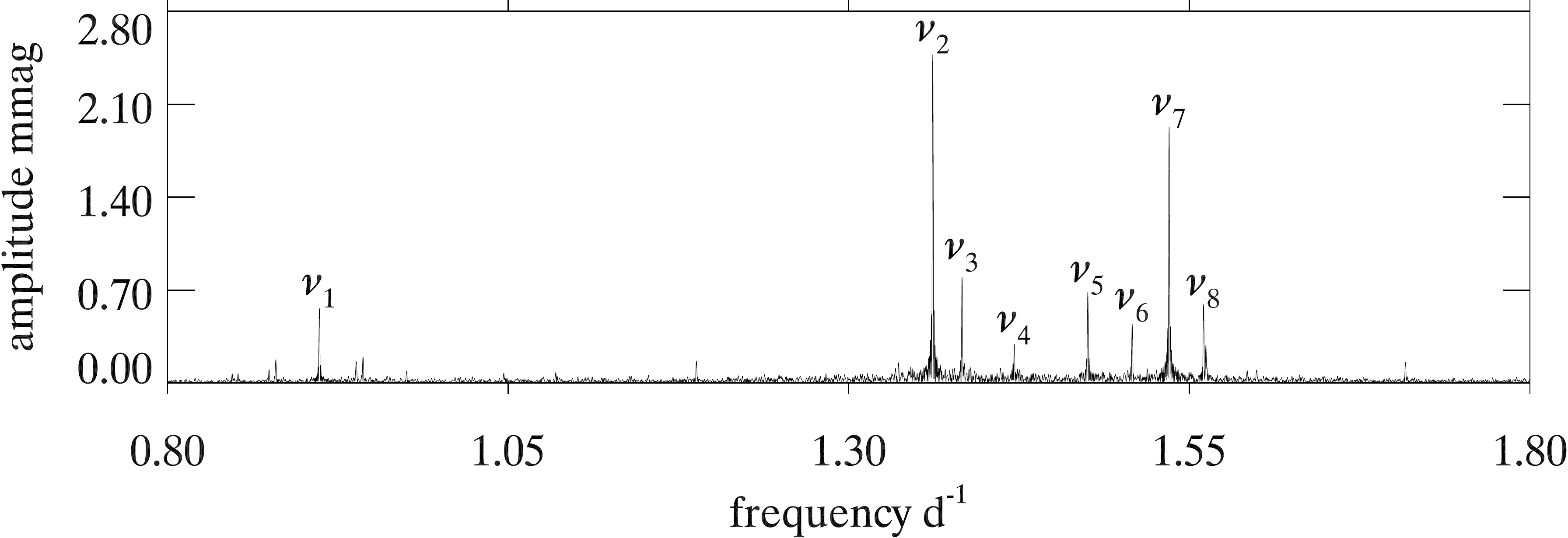}	
\caption{Left column: amplitude spectra for the $\gamma$~Dor star KIC\,7468196. Right column: amplitude spectra for the SPB star KIC\,10799291. Top panels: Amplitude spectra of the {\it Kepler} Q1-17 data out to 4\,d$^{-1}$ for both stars. There are no p-mode pulsations at higher frequencies up to the Nyquist frequency. Eye-catching features of these amplitude spectra are the frequency septuplets that make up the second frequency group, fg2, for both stars, but particularly for the SPB star on the right. These are consequences of combination frequencies of only three base frequencies. Middle panels: the amplitude spectra of the residuals after pre-whitening by the pulsation frequencies plus the combination frequencies of a subset of the base frequencies. These are given in Tables\,\ref{table:746} and \ref{table:1079}.  The scale is the same as in the top row, showing that almost all of the variance is explained for both stars. Bottom panels: Higher resolution looks at the fg1 frequency ranges with pulsation mode frequencies labelled. In the case of KIC\,10799291, only four of these were used as base frequencies to generate the combination frequencies, as shown in the tables. }
\label{fig:1079_746_ft}
\end{figure*}

As for $\gamma$~Dor stars such as KIC\,8113425 shown in Section\,\ref{sec:8113425}, many SPB stars also show frequency groups that have previously lacked an explanation (e.g., \citealt{mcnamaraetal2012}). In this section we compare the light curves and amplitude spectra for a $\gamma$~Dor star and an SPB star, showing a strong similarity that demonstrates that the SPB star light curve is also fully explained by a few base frequencies with combination frequencies, just as is the $\gamma$~Dor star light curve. The $\gamma$~Dor star, KIC\,7468196, has an upward light curve, and the SPB star, KIC\,10799291, has a downward light curve. We show that this difference in light curve shape arises only from the phases of the combination frequencies, and we conclude that this shows that the SPB star, with its downward light curve, is fully explained by a few g-mode pulsation frequencies and their combination frequencies. Fig.\,\ref{fig:1079_lc} compares typical 40-d sections from the light curves of the two stars.

Fig.\,\ref{fig:1079_746_ft} compares the amplitude spectra for the two stars, showing remarkable similarities. Both amplitude spectra are the result of a few base frequencies plus combination frequencies. For the $\gamma$~Dor star, KIC\,7468196, a frequency septuplet can be seen in the frequency range labelled as fg2 in the figure. Another striking frequency septuplet is seen for the SPB star, KIC\,10799291, also in the frequency range fg2. In both cases the central frequency of the fg2 septuplet is the sum of two of the base frequencies, and the other components are further combination frequencies. It is particularly notable that for KIC\,10799291 the amplitude of the highest peak in fg2, $\nu_2 + \nu_7$ in our notation, has a higher amplitude than either of its two base frequencies. For pulsating stars of the main-sequence it has not previously been recognised that the combination frequency amplitudes can be this large, hence they have usually not been recognised. We derived theoretically in Section\,\ref{sec:theory} that this is possible, and we demonstrate here observationally that it happens.

For the $\gamma$~Dor star KIC\,7468196 we fitted only three base frequencies and their combination frequencies with terms up to order $3\nu$, with the results shown in Table\,\ref{table:746}. As can be seen in the middle left panel of Fig.\,\ref{fig:1079_746_ft}, the reduction in the variance is striking. For the SPB star KIC\,10799291 there are more pulsation frequencies, and they are more widely spread in frequency, as can be seen in the right column of Fig.\,\ref{fig:1079_746_ft}. Because of their relatively high amplitudes, we fitted eight pulsation frequencies, but only used four of those as base frequencies for calculating the combination frequencies with terms up to order $2\nu$. The results are given in Table\,\ref{table:1079}.

\begin{table}
\centering
\caption[]{A least-squares fit of the three pulsation mode frequencies of the $\gamma$~Dor star KIC\,7468196 and the combination frequencies with terms up to order $3\nu$ for base frequencies $\nu_1$, $\nu_2$ and $\nu_3$ and with amplitudes restricted to greater than 0.3\,mmag. There are 27 identified frequencies, including the three base frequencies.  The zero point of the time scale is ${\rm BJD} \,245\,5694.25$}
\begin{tabular}{clrr}

\toprule
\multicolumn{1}{c}{labels} &
\multicolumn{1}{c}{frequency} & \multicolumn{1}{c}{amplitude} &
\multicolumn{1}{c}{phase} \\
&\multicolumn{1}{c}{d$^{-1}$} & \multicolumn{1}{c}{mmag} &
\multicolumn{1}{c}{radians}  \\
& & \multicolumn{1}{c}{$\pm 0.004$} &\\

\midrule
\multicolumn{4}{c}{fg0}\\
\midrule

$-\nu_1 + \nu_2   $ & $  0.1012839  $ & $  4.556    $ & $  0.8699  \pm  0.0009$ \\
$-2\nu_1 + 2\nu_2   $ & $  0.2025678  $ & $  1.890    $ & $  -0.4663  \pm 0.0021$ \\
$ -\nu_2 + \nu_3  $ & $  0.2337767  $ & $  1.190    $ & $  -0.9808  \pm  0.0034$ \\
$-3\nu_1 + 3\nu_2   $ & $  0.3038517  $ & $  0.678    $ & $  3.1209  \pm  0.0059$ \\
$-\nu_1 + \nu_3  $ & $  0.3350606  $ & $  1.993    $ & $  1.3341  \pm 0.0020$ \\
$-2\nu_1 + \nu_2 + \nu_3  $ & $  0.4363445  $ & $  0.606    $ & $  0.4556  \pm  0.0066$ \\

\midrule
\multicolumn{4}{c}{fg1}\\
\midrule

$ 3\nu_1 - 2\nu_2   $ & $  1.1632242  $ & $  0.688    $ & $  1.1930  \pm  0.0058$ \\
$  2\nu_2 - \nu_3  $ & $  1.2332992  $ & $  1.520    $ & $  2.7069  \pm 0.0026$ \\
$ 2\nu_1 - \nu_2   $ & $  1.2645081  $ & $  5.628    $ & $  -1.5188  \pm  0.0007$ \\
$-\nu_1 + 3\nu_2 - \nu_3  $ & $  1.3345831  $ & $  1.372    $ & $  1.8239  \pm 0.0029$ \\
$ \nu_1  $ & $  1.3657920 $ & $  23.879    $ & $  -2.1298  \pm  0.0002$ \\
$ 3\nu_1 - 3\nu_2 + \nu_3  $ & $  1.3970009  $ & $  0.333    $ & $  2.1111  \pm  0.0120$ \\
$  \nu_2  $ & $  1.4670759   $ & $  20.357    $ & $  0.6546  \pm  0.0002$ \\
$ 2\nu_1 - 2\nu_2 + \nu_3  $ & $  1.4982848  $ & $  0.392    $ & $  -0.2886  \pm  0.0102$ \\
$-\nu_1 + 2\nu_2   $ & $  1.5683598  $ & $  1.778    $ & $  -1.3550  \pm 0.0023$ \\
$-2\nu_1 + 3\nu_2   $ & $  1.6696437  $ & $  0.988    $ & $  -2.4776  \pm 0.0041$ \\
$  \nu_3  $ & $  1.7008526  $ & $  5.319    $ & $  1.8704  \pm  0.0008$ \\
$-\nu_1 + \nu_2 + \nu_3  $ & $  1.8021365  $ & $  0.558    $ & $  -0.1811  \pm  0.0072$ \\
$-2\nu_1 + 2\nu_2 + \nu_3  $ & $  1.9034204  $ & $  0.303    $ & $  -1.6220  \pm  0.0132$ \\

\midrule
\multicolumn{4}{c}{fg2}\\
\midrule

$ \nu_1 + 2\nu_2 - \nu_3  $ & $  2.5990912  $ & $  0.475    $ & $  -2.2965  \pm  0.0084$ \\
$ 3\nu_1 - \nu_2   $ & $  2.6303001  $ & $  1.519    $ & $  -0.2981  \pm 0.0026$ \\
$ 2\nu_1    $ & $  2.7315840  $ & $  2.066    $ & $  -0.8022  \pm  0.0019$ \\
$ \nu_1 + \nu_2   $ & $  2.8328679  $ & $  3.788    $ & $  1.9834  \pm  0.0011$ \\
$  2\nu_2   $ & $  2.9341518  $ & $  2.220    $ & $  -1.4248  \pm  0.0018$ \\
$ 2\nu_1 - \nu_2 + \nu_3  $ & $  2.9653607  $ & $  0.323    $ & $  -2.6336  \pm  0.0124$ \\
$ \nu_1 + \nu_3  $ & $  3.0666446  $ & $  0.953    $ & $  -3.0223  \pm 0.0042$ \\
$  \nu_2 + \nu_3  $ & $  3.1679285  $ & $  0.686    $ & $  0.0280  \pm  0.0058$ \\

\bottomrule

\end{tabular}
\label{table:746}
\end{table}

\begin{table}
\scriptsize
\centering
\caption[]{A least-squares fit of the eight pulsation mode frequencies of the SPB star KIC\,10799291 and the combination frequencies with terms up to order $2\nu$, for base frequencies $\nu_1$, $\nu_2$, $\nu_3$ and $\nu_7$, and with amplitudes restricted to greater than 50\,$\umu$mag. There are 41 identified frequencies, including the eight base frequencies.  The zero point of the time scale is ${\rm BJD} \,245\,5694.25$}
\begin{tabular}{clrr}
\toprule
\multicolumn{1}{c}{labels} &
\multicolumn{1}{c}{frequency} & \multicolumn{1}{c}{amplitude} &
\multicolumn{1}{c}{phase} \\
&\multicolumn{1}{c}{d$^{-1}$} & \multicolumn{1}{c}{mmag} &
\multicolumn{1}{c}{radians}  \\
& & \multicolumn{1}{c}{$\pm 0.004$} &\\

\midrule
\multicolumn{4}{c}{fg0}\\
\midrule

$2\nu_1 + \nu_3 - 2\nu_7$  &  $ 0.1357140$  &  $0.047 $  &  $-2.292 \pm 0.093 $ \\
$-\nu_1+ \nu_2 + 2\nu_3 - 2\nu_7$  &  $ 0.1462140$  &  $0.096 $  &  $2.126 \pm 0.045 $ \\
$-\nu_3 + \nu_7$  &  $ 0.1520070$  &  $0.221 $  &  $1.719 \pm 0.020 $ \\
$-\nu_2 + \nu_7$  &  $ 0.1734908$  &  $1.297 $  &  $-0.322 \pm 0.003 $ \\
$-\nu_1 + 2\nu_3 - \nu_7$  &  $ 0.3197048$  &  $0.039 $  &  $1.992 \pm 0.112 $ \\
$-2\nu_2 + 2\nu_7$  &  $ 0.3469816$  &  $0.106 $  &  $-1.748 \pm 0.041 $ \\
$-\nu_1+ 2\nu_2 - \nu_3$  &  $0.4287442$  &  $0.053 $  &  $2.803 \pm 0.082 $ \\
$-\nu_1+ \nu_2$  &  $0.4502280$  &  $0.274 $  &  $1.069 \pm 0.016 $ \\
 $2\nu_1 - 2\nu_3 + \nu_7$  &  $ 0.5917350$  &  $0.043 $  &  $-0.009 \pm 0.101 $ \\
$-\nu_1+ \nu_2 - \nu_3 + \nu_7$  &  $ 0.6022350$  &  $0.263 $  &  $0.251 \pm 0.017 $ \\
$-\nu_1+ \nu_7$  &  $ 0.6237188$  &  $0.426 $  &  $0.909 \pm 0.010 $ \\
 $\nu_1+ \nu_2 - \nu_7$  &  $ 0.7379490$  &  $0.049 $  &  $-0.582 \pm 0.088 $ \\

\midrule
\multicolumn{4}{c}{fg1}\\
\midrule

 $\nu_1$ &  $0.9114398 $  &  $0.565 $  &  $-1.668 \pm 0.008 $ \\
$-2\nu_1 + 2\nu_3$  &  $0.9434236$  &  $0.188 $  &  $-1.837 \pm 0.023 $ \\
 $\nu_1- \nu_2 + \nu_7$  &  $ 1.0849306$  &  $0.082 $  &  $-1.998 \pm 0.053 $ \\
 $2\nu_2 + \nu_7$  &  $ 1.1881770$  &  $0.160 $  &  $-0.889 \pm 0.027 $ \\
 $\nu_2$  &  $1.3616678 $  &  $2.474 $  &  $-1.487  \pm 0.002$ \\
 $\nu_3$  &  $1.3831516 $  &  $0.841 $  &  $-3.141 \pm 0.005 $ \\
$\nu_4$  &  $1.4214710 $  &  $0.286 $  &  $0.616 \pm 0.015 $ \\
$\nu_5$  &  $1.4754643 $  &  $0.671 $  &  $-0.629 \pm 0.007 $ \\
 $2\nu_1 + 2\nu_2 - 2\nu_7$  &  $ 1.4758980$  &  $0.022 $  &  $2.876 \pm 0.222 $ \\
$-\nu_1+ 2\nu_2 + 2\nu_3 - 2\nu_7$  &  $ 1.5078818$  &  $0.111 $  &  $0.317 \pm 0.067 $ \\
$\nu_6$  &  $1.5081214 $  &  $0.449 $  &  $1.902 \pm 0.017 $ \\
 $\nu_7$  &  $ 1.5351586 $  &  $1.927 $  &  $-1.626  \pm 0.002$ \\
$\nu_8$  &  $1.5604683 $  &  $0.568 $  &  $-1.683 \pm 0.008 $ \\

\midrule
\multicolumn{4}{c}{fg2}\\
\midrule

 $-\nu_2 + 2\nu_7$  &  $ 1.7086494$  &  $0.160 $  &  $-1.929 \pm 0.027 $ \\
$-\nu_1+ 2\nu_2$  &  $1.8118958$  &  $0.083 $  &  $-0.409 \pm 0.053 $ \\
$-\nu_1+ \nu_2 + \nu_7$  &  $ 1.9853866$  &  $0.135 $  &  $-0.672 \pm 0.032 $ \\
$-\nu_1+ 2\nu_7$  &  $ 2.1588774$  &  $0.094 $  &  $-1.110 \pm 0.046 $ \\
 $\nu_1+ \nu_2$  &  $2.2731076$  &  $0.294 $  &  $-3.055 \pm 0.015 $ \\
 $\nu_1+ \nu_7$  &  $ 2.4465984$  &  $0.322 $  &  $2.872 \pm 0.014 $ \\
 $2\nu_2$  &  $2.7233356$  &  $1.244 $  &  $-3.020  \pm 0.004$ \\
 $\nu_2 + \nu_3$  &  $2.7448194$  &  $0.096 $  &  $1.771 \pm 0.045 $ \\
 $\nu_2 + \nu_7$  &  $ 2.8968264$  &  $2.492 $  &  $2.985  \pm 0.002$ \\
 $\nu_3 + \nu_7$  &  $ 2.9183102$  &  $0.081 $  &  $1.707 \pm 0.054 $ \\
$-\nu_1+ 2\nu_2 + 2\nu_3 - \nu_7$  &  $ 3.0430404$  &  $0.058 $  &  $-1.096 \pm 0.074 $ \\
 $2\nu_7$  &  $ 3.0703172$  &  $1.184 $  &  $2.628  \pm 0.004$ \\
$-\nu_1+ 2\nu_2 + \nu_7$  &  $ 3.3470544$  &  $0.430 $  &  $-2.237 \pm 0.010 $ \\
$-\nu_1+ \nu_2 + 2\nu_7$  &  $ 3.5205452$  &  $0.314 $  &  $-2.588 \pm 0.014 $ \\
 $2\nu_2 + \nu_7$  &  $ 4.2584942$  &  $0.064 $  &  $2.436 \pm 0.068 $ \\
 $\nu_2 + 2\nu_7$  &  $ 4.4319850$  &  $0.076 $  &  $2.071 \pm 0.057 $ \\

\bottomrule

\end{tabular}
\label{table:1079}
\end{table}

\subsection{The phases}

\begin{figure}
\centering	
\includegraphics[width=0.9\linewidth,angle=0]{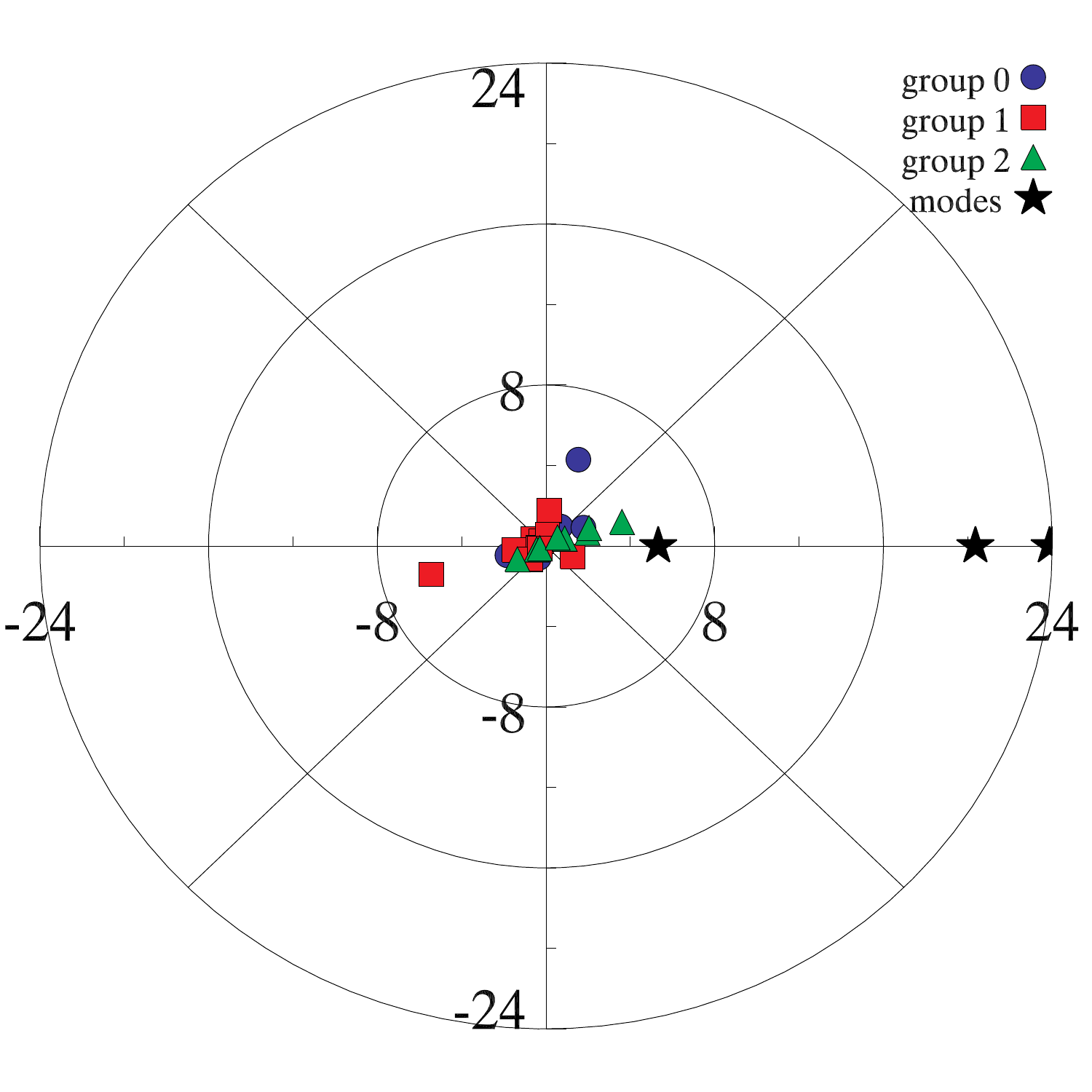}	
\includegraphics[width=0.9\linewidth,angle=0]{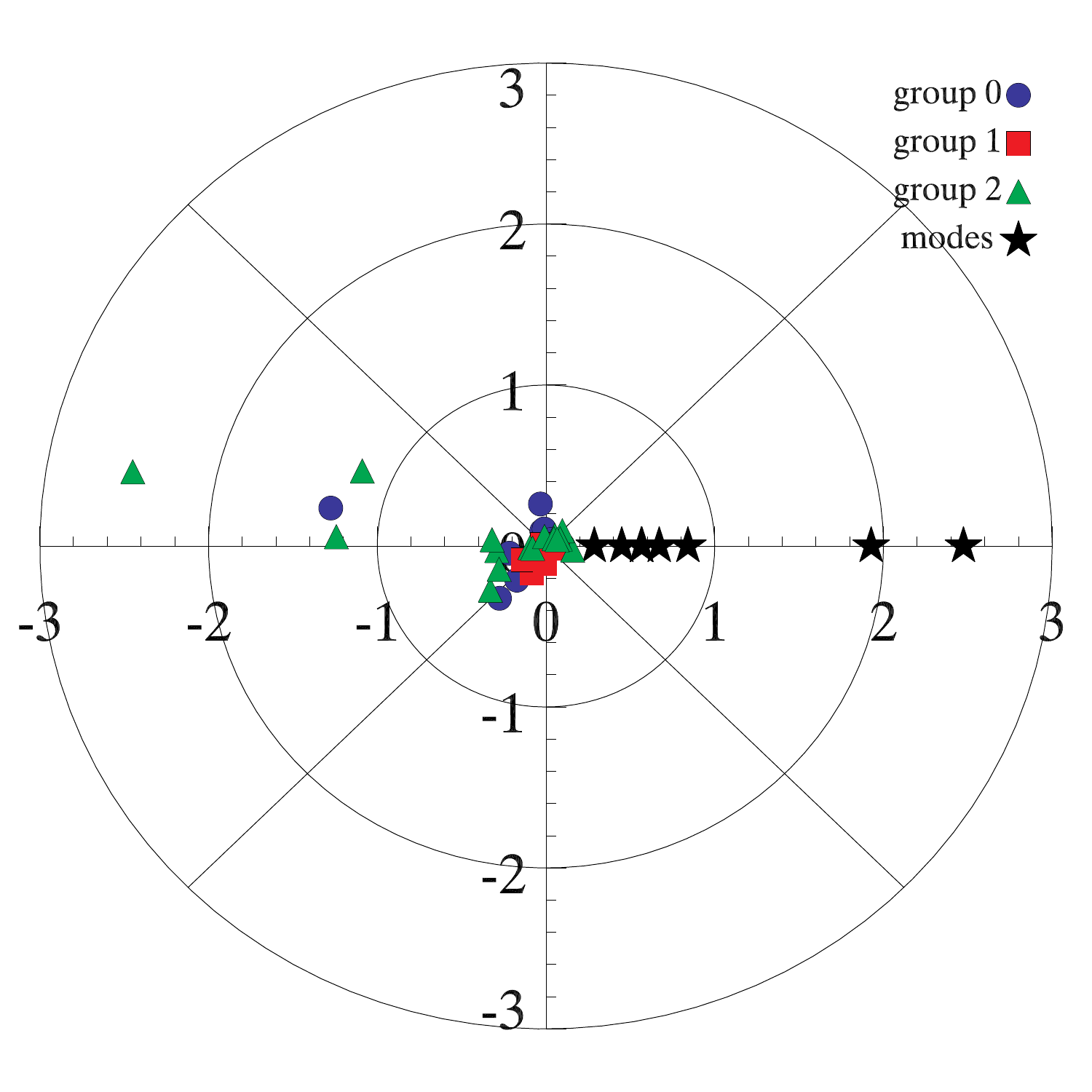}	
\caption{Phasor plots for for the $\gamma$~Dor star KIC\,7468196 (top) and for the SPB star KIC\,10799291 (bottom). It can be seen that the difference between the light curves of these two stars (Fig.\,\ref{fig:1079_lc}) -- upward for the $\gamma$~Dor star and downward for the SPB star -- is a consequence of the phases of the combination frequencies.
}
\label{fig:746_1079_phasor}
\end{figure}

The contrasting shapes of the light-curve asymmetries in KIC\,7468196 and KIC\,10799291 lead to contrasting phasor plots, as shown in Fig.\,\ref{fig:746_1079_phasor}. The combination frequencies belonging to KIC\,10799291 lie almost entirely on the left of the diagram, describing strong, downward asymmetry (bottom panel). Conversely, the distribution of points in the upper panel is not strongly skewed to one side, indicating that the asymmetry in the light curve is not strong. Indeed, close inspection of the top panel of Fig.\,\ref{fig:1079_lc} shows that this is the case -- the light curve has only mild upward asymmetry. The fact that we observe asymmetry at all, when the distribution is so even, is an effect of the frequencies of the combinations. Those that lie in fg2 have higher frequencies, and thus simultaneously suppress the minima while reinforcing the maxima. The small upward asymmetry is thus explained by the small imbalance of the fg2 frequencies to the right of the diagram.

\subsection{Discussion of the comparison of KIC\,7468196 and KIC\,10799291}

We argue from the similarities of the amplitude spectra of the $\gamma$~Dor star KIC\,7468196 and the SPB star KIC\,10799291, and from the fact that most of the variance in both light curves is explained by a few base frequencies and their combination frequencies, that both stars are pulsating in g~modes that fully account for their light variations. While there are further details in the amplitude spectra at lower amplitudes that we have not extracted, we expect that these details will be accounted for by pulsation in other modes and the additional combination frequencies that these may generate. The difference in the appearance of the light curves, with the $\gamma$\,Dor light curve being upward and the SPB star light curve being downward, is a consequence of the phases of the combination frequencies. There are $\gamma$~Dor stars with downward light curves, too. There is no need to conjecture spots to explain this, and spots do not generate combination frequencies, hence are not a viable hypothesis. The light curves of the $\gamma$~Dor and SPB stars are fully explained by g~mode pulsation. We will expand on this point in the final discussion.

There has been some confusion about the classification of KIC\,10799291 that we clarify here. KIC\,10799291 has revised {\it Kepler} parameters $T_{\rm eff} = 11\,100 \pm 400$~K and  $\log g =4.4 \pm 0.1$ (cgs units) \citep{huberetal2014}, as listed in Table\,\ref{tab:data}. However, the original KIC parameters are $T_{\rm eff} = 10\,950$~K and $\log g = 6.1$, which suggested a white dwarf star. \citet{mcnamaraetal2012}  also classified this star as a white dwarf based on proper motion and the original KIC parameters. That classification propagated into SIMBAD, where KIC\,10799291 is listed as a `pulsating white dwarf'. The revised KIC parameters of \citet{huberetal2014} and the frequencies of the variation in the star show that it is an SPB star, not a white dwarf. 

\clearpage
\clearpage

\section{Two more SPB stars}
\label{sec:twomore}

To emphasise the point that the SPB stars are purely g-mode pulsators, and to extend the application of combination frequencies to understanding these stars, we show two further cases where we do not agree with the classification of the star in the literature: KIC\,10118750, which has a downward light curve and has been suggested to be a $\beta$\,Cep star, and KIC\,5450881, which has a light curve that would probably have been called `symmetric' in previous discussions and which has been previously classified as a rotational or orbital variable. In both cases we view these stars as SPB stars with combination frequencies. 

\begin{figure}
\centering	
\includegraphics[width=0.99\linewidth,angle=0]{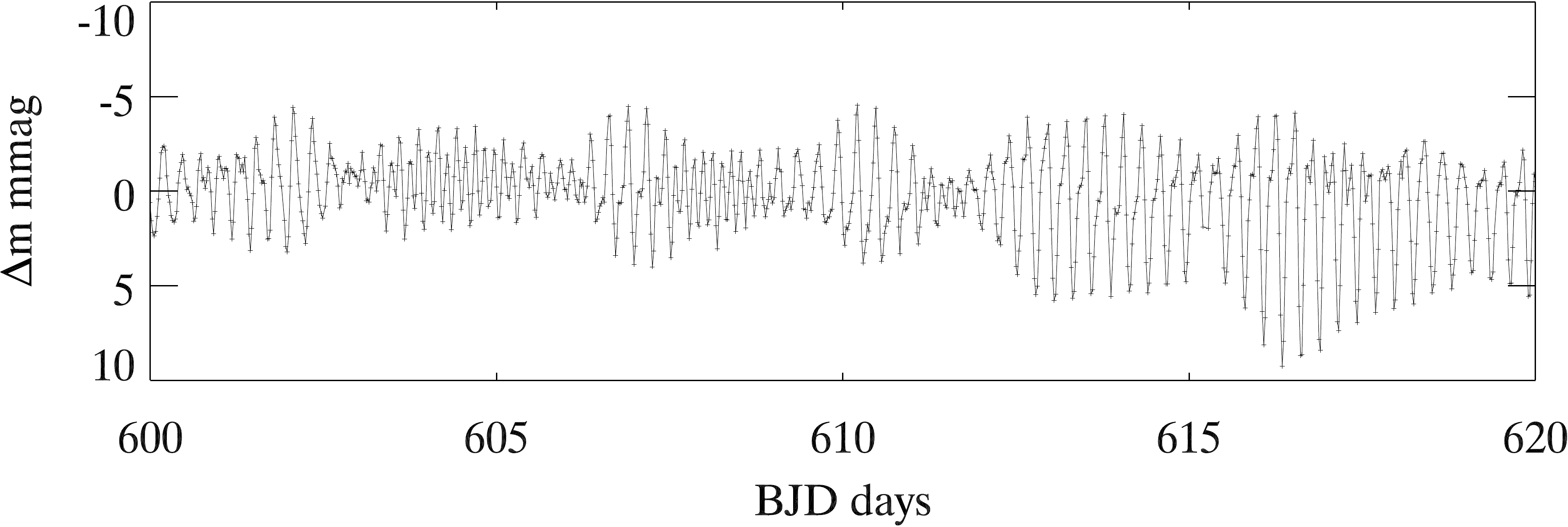}	
\caption{A section of the light curve of KIC\,10118750 spanning 20~d showing the symmetric, then downward character of the light variations. The time is relative to ${\rm BJD} \,245\,5000$. }
\label{fig:1011_lc}
\end{figure}

\subsection{KIC\,10118750}
\label{sec:1011}

Fig.\,\ref{fig:1011_lc} shows a section of the light curve of KIC\,10118750 where the at first symmetric, then downward appearance can be seen. \citet{mcnamaraetal2012} classified this star as a $\beta$\,Cep star, probably because of the higher frequency range and frequency groups compared to some SPB stars. We find that there are only five pulsation modes, all with frequencies in the $3.6 -4.2$\,d$^{-1}$ range, which could be consistent with a $\beta$\,Cep classification. However, the combination frequencies, the downward light curve and, especially, the late-B effective temperature, $T_{\rm eff} = 11400$\,K, suggest instead that this is an SPB star.

\begin{figure}
\centering	
\includegraphics[width=0.99\linewidth,angle=0]{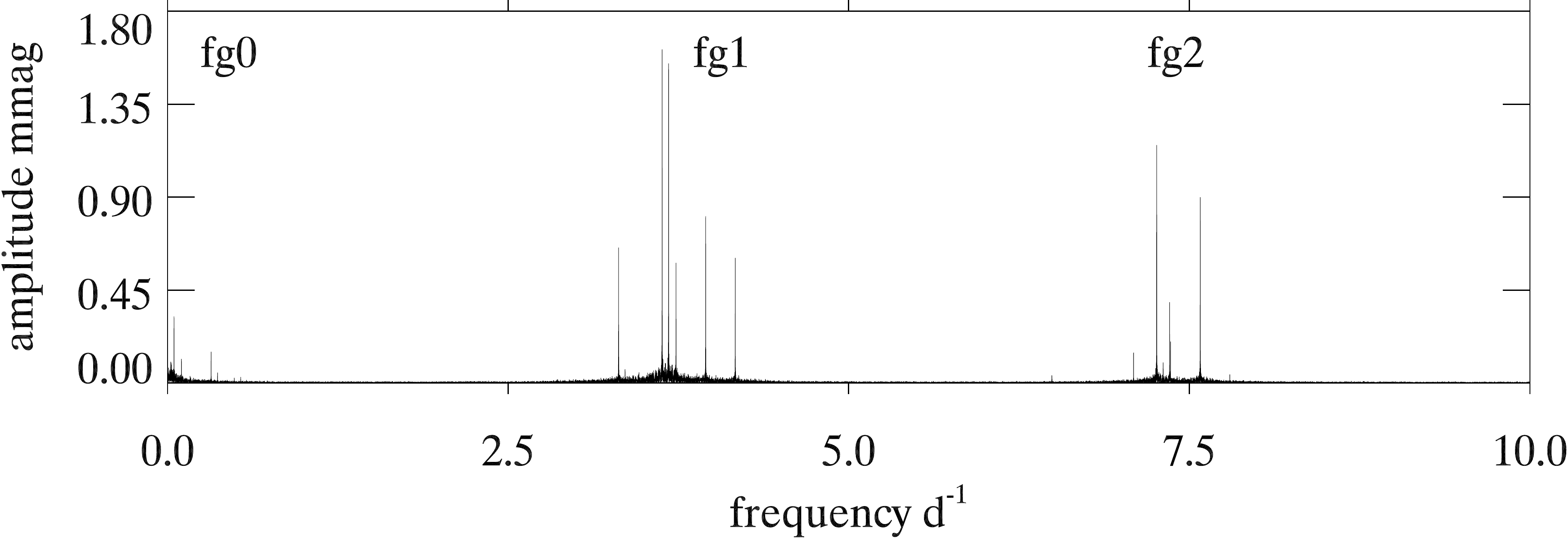}	
\includegraphics[width=0.99\linewidth,angle=0]{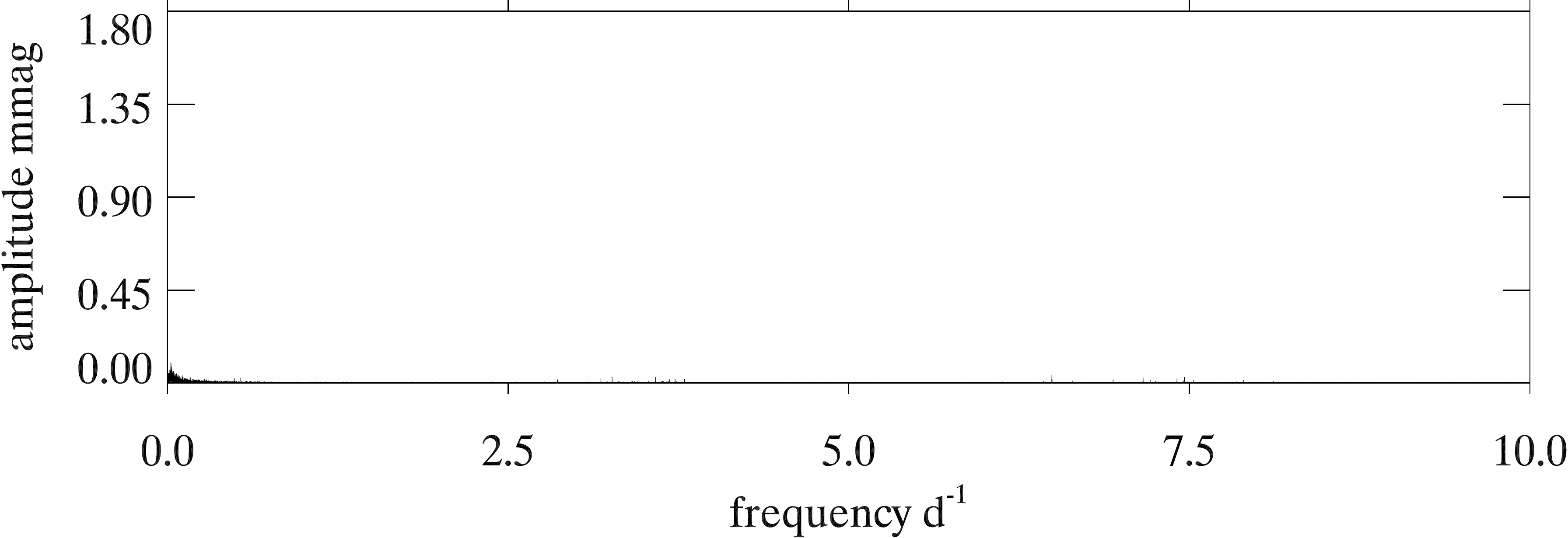}	
\includegraphics[width=0.99\linewidth,angle=0]{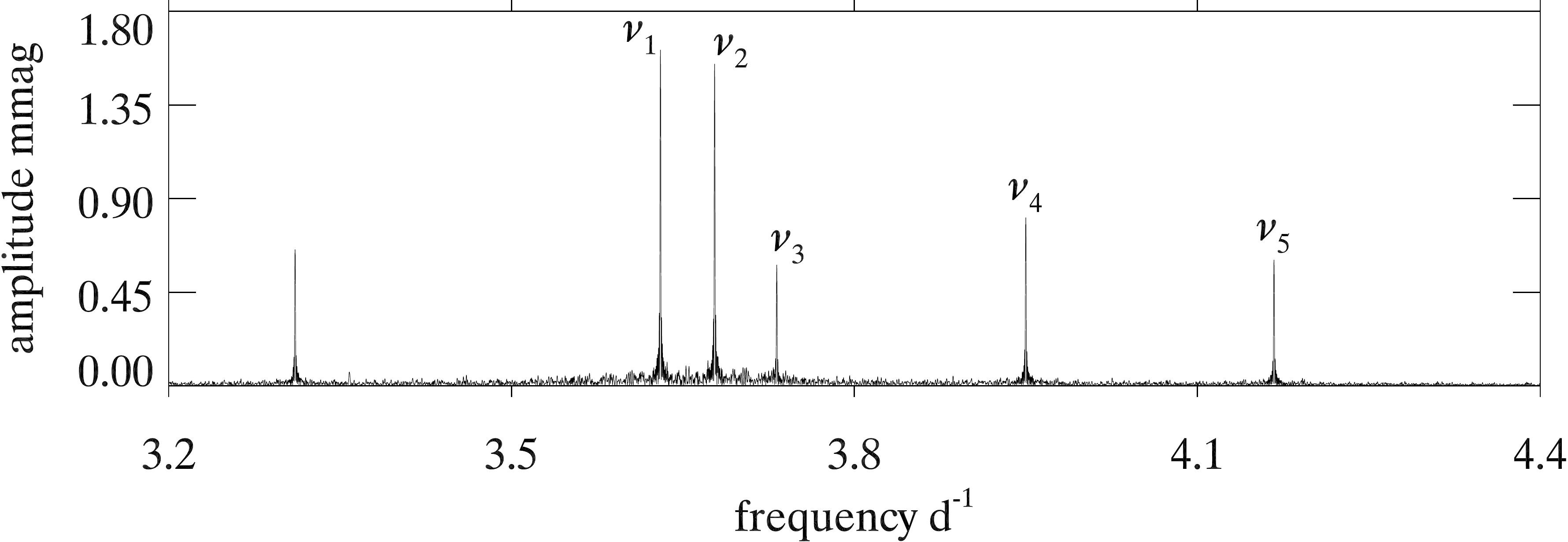}	
\caption{Top panel: An amplitude spectrum for KIC\,10118750 out to 10\,d$^{-1}$. There are no p~mode pulsations; the higher frequency peaks are combination frequencies from five pulsation mode frequencies in fg1. Because the higher frequency groups are composed of higher order combination frequencies, we concentrate on the groups fg0, fg1 and fg2 here to show the result. Middle panel: an amplitude spectrum of the residuals after fitting the five pulsation frequencies and 26 combination frequencies with amplitudes above 30\,$\umu$mag. There are significant peaks remaining with amplitudes below our imposed 30\,$\umu$mag limit; further combination frequencies probably explain those, too. The ordinate scale has been kept the same in all panels to give the impact of the variance reduction with such a simple explanation.  Bottom panel: An expanded amplitude spectrum of the {\it Kepler} Q1-17 data for KIC\,10799291 out to 10\,d$^{-1}$ with the pulsation frequencies $\nu_1$ to $\nu_5$ marked. Note that the lowest unmarked peak in fg1 could be substituted for $\nu_4$ and the same solution would be found. }
\label{fig:ft_1011}
\end{figure}

KIC\,10118750 has five principal base frequencies that we attribute to g-mode pulsations; most of the rest of the variance is in the combination frequencies. Fig.\,\ref{fig:ft_1011} shows the amplitude spectrum with frequency groups fg0, fg1 and fg2 marked in the top panel. There are higher frequency groups of much lower amplitude that can be seen, but are not shown here. The bottom panel expands the frequency range around fg1 and labels the five base frequencies selected. Those five frequencies and their combination frequencies with terms up to order $2\nu$ were fitted by least-squares to the Q0--17 data set. We restricted the fit to the five base frequencies and only those combination frequencies that had amplitudes greater than 30\,$\umu$mag in the frequency range up to 10\,d$^{-1}$, resulting in 31 frequencies given in Table\,\ref{table:1011}. 

The middle panel of Fig.\,\ref{fig:ft_1011} shows the stunning removal of all peaks in fg0, fg1 and fg2 when the 31 frequencies from Table\,\ref{table:1011} are fitted to the data. Even the very low amplitude peaks remaining can be modelled with additional combination frequencies\footnote{We chose to keep the number of fitted frequencies low for the impact this has. It is not our purpose here to generate thousands of combination frequencies to explain all of the tiny variance remaining.}. We conclude that KIC\,10118750 is an SPB star with a relatively high g-mode frequency range for the class.

\begin{table}
\scriptsize
\centering
\caption[]{A least-squares fit of the five base frequencies of KIC\,10118750 and the combination frequencies with terms up to order $2\nu$, with frequencies up to 10\,d$^{-1}$, and with amplitudes restricted to greater than 30\,$\umu$mag. There are 31 identified frequencies, including the five base frequencies.  The zero point of the time scale is ${\rm BJD} \,245\,5694.25$. }
\begin{tabular}{clrr}

\toprule

\multicolumn{1}{c}{labels} &
\multicolumn{1}{c}{frequency} & \multicolumn{1}{c}{amplitude} &
\multicolumn{1}{c}{phase} \\
&\multicolumn{1}{c}{d$^{-1}$} & \multicolumn{1}{c}{mmag} &
\multicolumn{1}{c}{radians}  \\
& & \multicolumn{1}{c}{$\pm 0.002$} &\\

\midrule
\multicolumn{4}{c}{fg0}\\
\midrule

$\nu_1 -2\nu_2 + \nu_3   $ & $ 0.0069965 $ & $0.038 $ & $2.279 \pm 0.059 $ \\
$ \nu_1 -2\nu_2  + 2\nu_4 -\nu_5 $ & $ 0.0077650 $ & $0.066 $ & $1.692 \pm 0.033 $ \\
$ 2\nu_1  -2\nu_3 -\nu_4 + \nu_5 $ & $ 0.0136230 $ & $0.032 $ & $-1.947 \pm 0.070 $ \\
$-\nu_1 + \nu_2    $ & $ 0.0473818 $ & $0.309 $ & $-2.138 \pm 0.007 $ \\
$-\nu_1 + \nu_2 -\nu_3 + 2\nu_4 -\nu_5 $ & $ 0.0481503 $ & $0.026 $ & $-2.966 \pm 0.087 $ \\
$  -\nu_2 + \nu_3   $ & $ 0.0543783 $ & $0.046 $ & $0.403 \pm 0.048 $ \\
$ \nu_1 + \nu_2 -2\nu_3 -\nu_4 + \nu_5 $ & $ 0.0610048 $ & $0.040 $ & $-1.670 \pm 0.056 $ \\
$-2\nu_1 + 2\nu_2    $ & $ 0.0947636 $ & $0.038 $ & $2.440 \pm 0.059 $ \\
$-\nu_1  + \nu_3   $ & $ 0.1017601 $ & $0.113 $ & $-1.050 \pm 0.020 $ \\
$-\nu_1   + \nu_4  $ & $ 0.3196718 $ & $0.155 $ & $-1.740 \pm 0.014 $ \\
$-2\nu_1 + \nu_2  + \nu_4  $ & $ 0.3670536 $ & $0.052 $ & $2.598 \pm 0.042 $ \\

\midrule
\multicolumn{4}{c}{fg1}\\
\midrule

$ 2\nu_1   -\nu_4  $ & $ 3.3105639 $ & $0.655 $ & $-1.634 \pm 0.003 $ \\
$ 2\nu_1  -\nu_3 + \nu_4 -\nu_5 $ & $ 3.3113324 $ & $0.030 $ & $-2.996 \pm 0.075 $ \\
$ \nu_1 + \nu_2  -\nu_4  $ & $ 3.3579457 $ & $0.058 $ & $-0.470 \pm 0.038 $ \\
$ \nu_1 + \nu_2 -\nu_3 + \nu_4 -\nu_5 $ & $ 3.3587142 $ & $0.034 $ & $-1.589 \pm 0.066 $ \\
$ \nu_2  + \nu_4 -\nu_5 $ & $ 3.4604743 $ & $0.066 $ & $0.647 \pm 0.034$ \\
$ \nu_1     $ & $ 3.6302357$ & $1.645 $ & $-1.213 \pm 0.001 $ \\
$ \nu_2    $ & $ 3.6776175 $ & $1.581 $ & $-2.979 \pm 0.001 $ \\
$ \nu_3   $ & $ 3.7319958 $ & $0.570 $ & $-1.726 \pm 0.004 $ \\
$ \nu_1 -\nu_2  + \nu_4  $ & $ 3.9025257$ & $0.049 $ & $-0.597 \pm 0.045 $ \\
$  \nu_4  $ & $ 3.9499075 $ & $0.796 $ & $-2.697 \pm 0.003 $ \\
$  \nu_5 $ & $ 4.1670507 $ & $0.601 $ & $-2.628  \pm 0.004$ \\

\midrule
\multicolumn{4}{c}{fg2}\\
\midrule

$ \nu_1 + \nu_2  + \nu_4 -\nu_5 $ & $ 7.0907100 $ & $0.141 $ & $2.409 \pm 0.016 $ \\
$ \nu_1 + 2\nu_2 -\nu_3   $ & $ 7.2534749 $ & $0.051 $ & $-1.693 \pm 0.043 $ \\
$ 2\nu_1     $ & $ 7.2604714 $ & $1.151 $ & $-1.624  \pm 0.002$ \\
$ \nu_1 + \nu_2    $ & $ 7.3078532 $ & $0.131 $ & $2.503 \pm 0.017 $ \\
$ 2\nu_2    $ & $ 7.3552350 $ & $0.401 $ & $0.513 \pm 0.006 $ \\
$ \nu_1  + \nu_3   $ & $ 7.3622315 $ & $0.196 $ & $-2.437 \pm 0.011 $ \\
$ \nu_1   + \nu_4  $ & $ 7.5801432 $ & $0.894 $ & $2.767 \pm 0.003 $ \\
$ \nu_2  + \nu_4  $ & $ 7.6275250 $ & $0.047 $ & $0.670 \pm 0.047 $ \\
$ \nu_1    + \nu_5 $ & $ 7.7972864 $ & $0.036 $ & $2.864 \pm 0.062 $ \\

\bottomrule

\end{tabular}
\label{table:1011}
\end{table}

\subsection{The phases}

\begin{figure}
\centering	
\includegraphics[width=0.9\linewidth,angle=0]{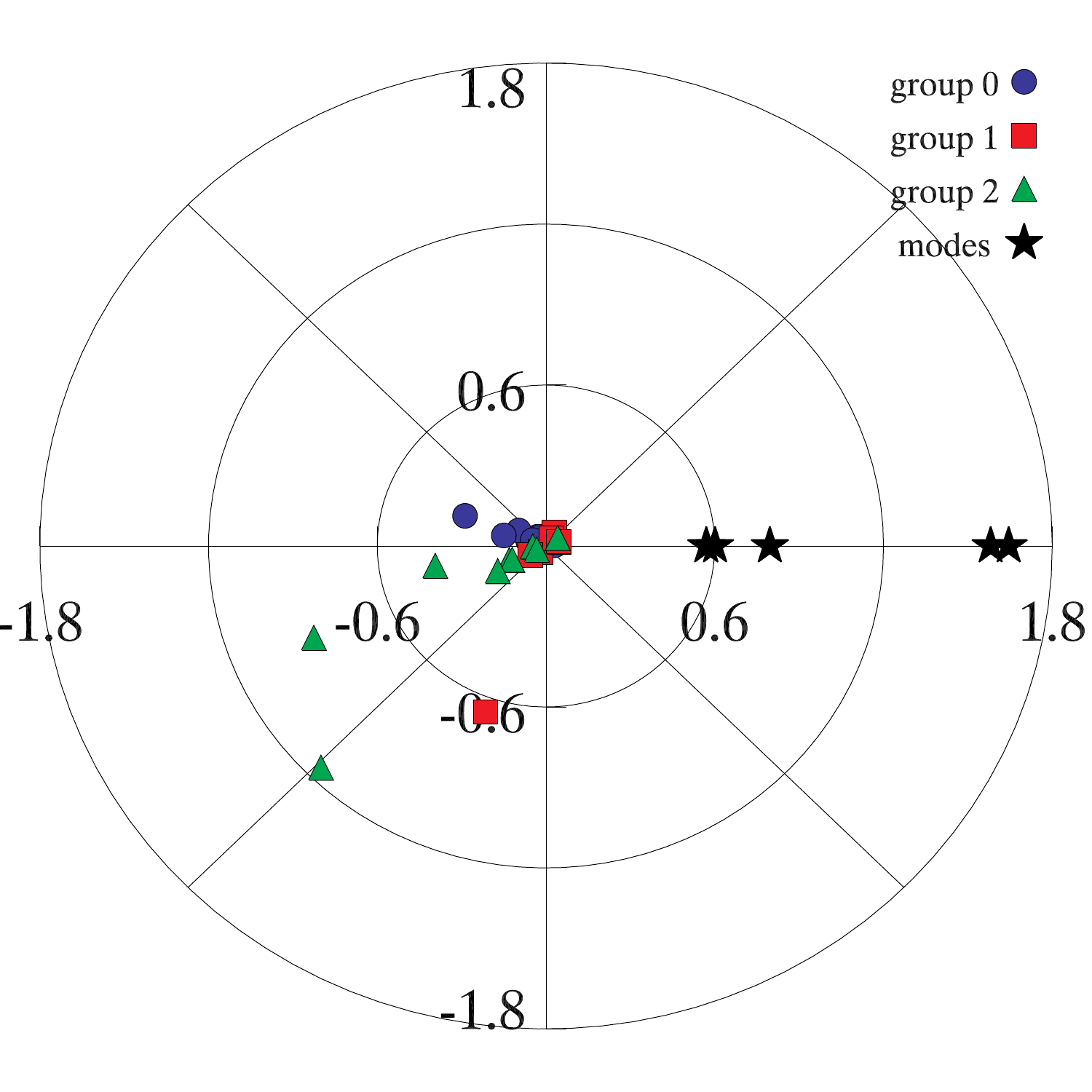}
\caption{The phasor plot for the SPB star KIC\,10118750.}
\label{fig:1011_phasor}
\end{figure}

The strong downward light curve is described by combination frequencies with relative phases near $\upi$. The minima in this star reach abnormal depths, even for stars with asymmetric light curves, as is seen to the right of Fig.\,\ref{fig:1011_lc}. As such, the phasor plot in Fig.\,\ref{fig:1011_phasor} shows points that not only lie predominantly on the left, but that also have amplitudes of similar magnitude to their base frequencies.

\clearpage
\clearpage

\subsection{KIC\,5450881 -- the simplest case of a frequency group star}
\label{sec:545}

\begin{figure}
\centering	
\includegraphics[width=0.99\linewidth,angle=0]{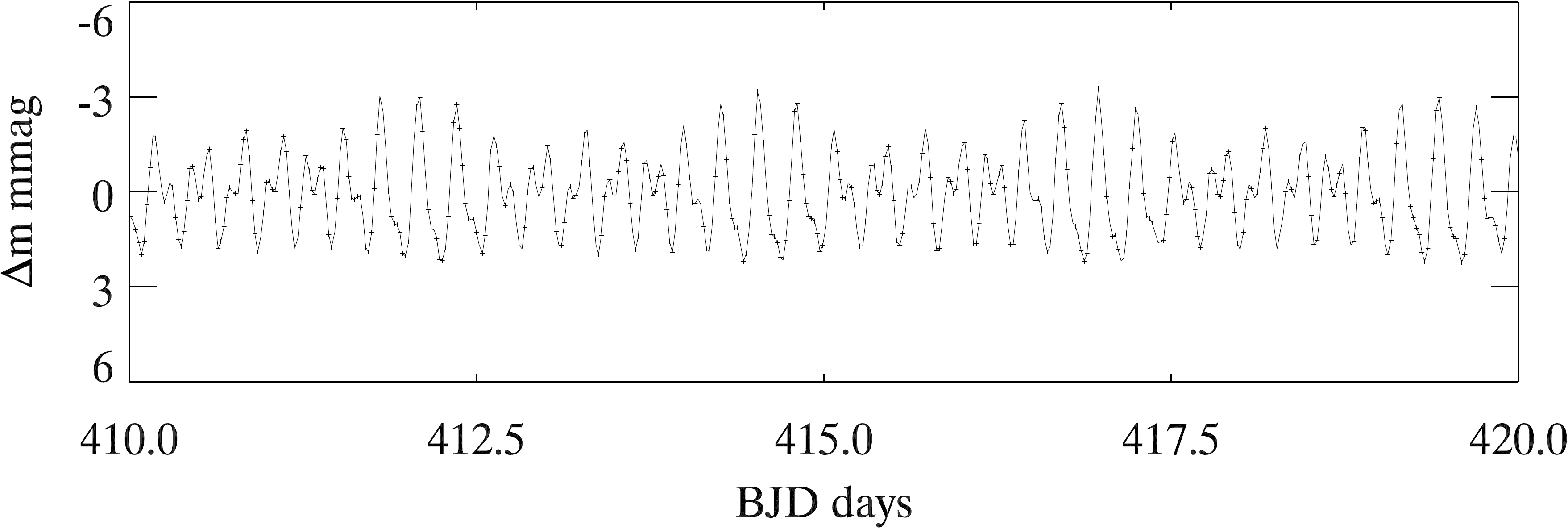}	
\caption{A section of the light curve of KIC\,5450881 spanning 10~d. The time is relative to ${\rm BJD}\,245\,5000$. Almost all of the variation in the light curve is explained by two base frequencies and six combination frequencies. A rotational hypothesis with spots is not viable. This is pulsation.}
\label{fig:545_lc}
\end{figure}

\begin{figure}
\centering	
\includegraphics[width=0.99\linewidth,angle=0]{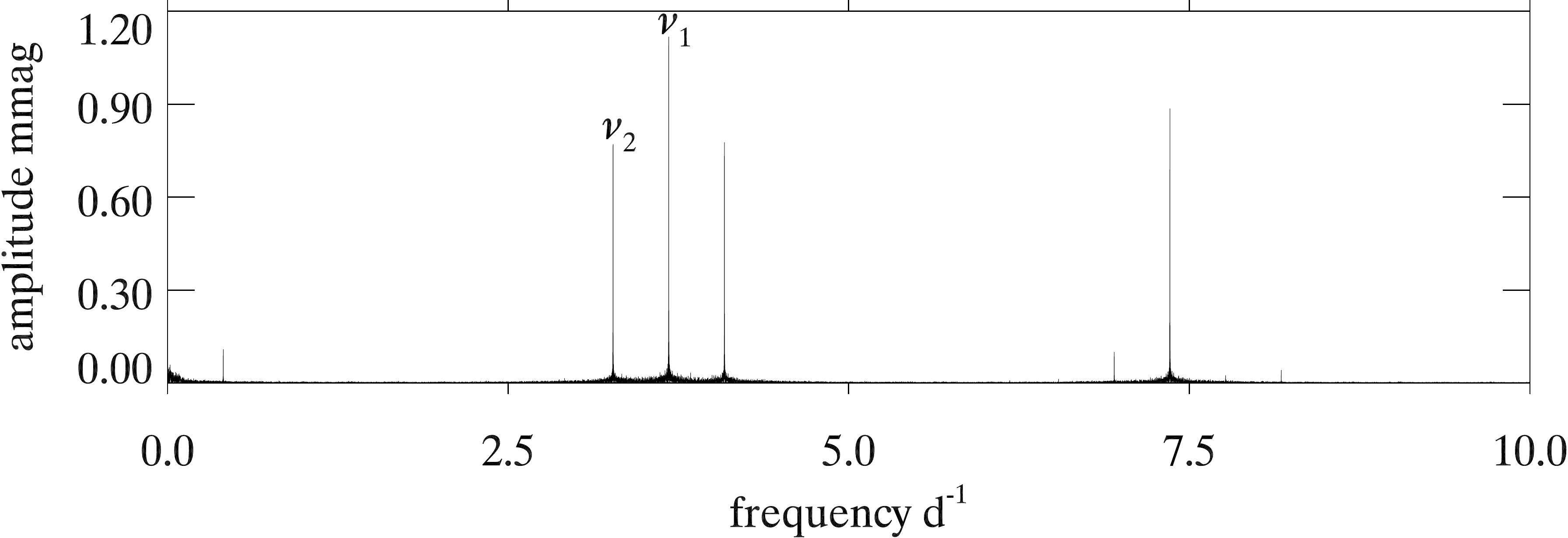}	
\includegraphics[width=0.99\linewidth,angle=0]{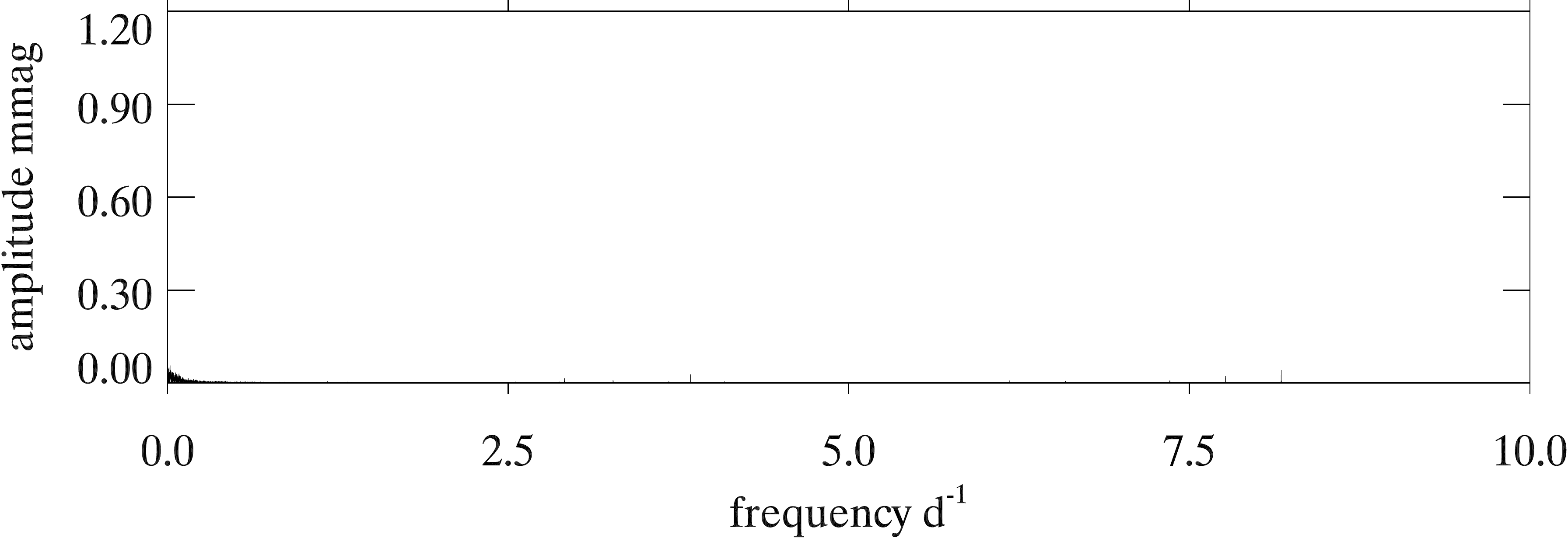}	
\caption{Top panel: An amplitude spectrum for KIC\,5450881 out to 10\,d$^{-1}$ showing a triplet and combination frequencies. The two marked peaks were chosen as the base frequencies to generate the combination frequencies; the other member of the triplet would serve equally well. Bottom panel: the amplitude spectrum of the residuals after pre-whitening by the two base frequencies and six combination frequencies with terms up to order $2\nu$.  }
\label{fig:ft_545}
\end{figure}

\citet{mcnamaraetal2012} listed KIC\,5450881 as either a rotational or binary variable. Fig.~\ref{fig:545_lc} shows a 10-d section of the light curve, which can be described with two base frequencies and six combination frequencies. This can be seen in Fig.\,\ref{fig:ft_545} where a dominant triplet and combination frequencies are evident. Because the triplet is equally spaced, we only need two base frequencies; we have arbitrarily chosen the two labelled. We then find only six combination frequencies, one of which is the other member of the triplet. With the two base frequencies and six combination frequencies, almost all of the variance is strikingly removed, as can be seen in the lower panel of Fig.\,\ref{fig:ft_545}. At higher amplitude resolution additional significant peaks can be seen in this frequency range and at higher frequency; these can be modelled in a more detailed analysis in the future.

This pattern of two base frequencies, which we propose are from two g~modes,  and six combination frequencies gives an elegant explanation of the light curve that is preferable to a rotational spot model or a binary star model. Spots and binary variations do not give rise to combination frequencies. This star is thus the simplest case of frequency group star, with only two base frequencies. The other stars presented in this paper with more base frequencies are more complex examples of the same physics, hence this star illustrates the principles most clearly. 

The equally spaced frequency triplet could have one of several origins. It could be interpreted as a rotational triplet, hence probably arising from a g-mode dipole. Alternatively, it could arise from pure geometric amplitude modulation, such as is seen in roAp stars, or pure frequency modulation, as could arise if there were a very massive companion \citep{fm2012}. Both of the latter possibilities can be tested. We chose a time, $t_0$, that set the phases equal for the two sidelobes to the central peak. With two sinusoidal variations this is always possible. We then expect for pure amplitude modulation that the central frequency should have the same phase at $t_0$, and for pure frequency modulation the central frequency should have a phase $\upi/2$ different to the sidelobe phase. 

Neither of these conditions is obtained in this case -- the phase difference between the central frequency and the sidelobes is 2.2\,rad -- hence we conclude that the triplet arises neither from pure amplitude modulation, nor from pure frequency modulation. It can then be represented with a rotational triplet, or simply two base g-mode frequencies with combination frequencies.  If the triplet is from rotationally split dipole modes of of order $m = -1, 0, +1$, then the the inverse splitting of the triplet is $1/(\nu_2 - \nu_1) = 2.4466$\,d, and an estimate of the Ledoux constant leads to a rotational period for the star somewhat less than twice that, i.e., $\sim 4-5$\,d. A spectroscopic measurement of $v \sin i$ could constrain or refute that possibility. \citet{bregerkolenberg2006} examined several cases of nearly equally split triplets in a variety of pulsating stars and suggested that in those cases the triplets were better understood in terms of two base mode frequencies and combination frequencies. That may be the case, also, for KIC\,5450881. 

\begin{table}
\centering
\caption[]{A least-squares fit of the two base frequencies of KIC\,5450881 and six combination frequencies with terms up to order $2\nu$, where one of the combinations is the other member of the dominant triplet.  The zero point of the time scale is ${\rm BJD} \,245\,5694.25$}
\begin{tabular}{clrr}

\toprule

\multicolumn{1}{c}{labels} &
\multicolumn{1}{c}{frequency} & \multicolumn{1}{c}{amplitude} &
\multicolumn{1}{c}{phase} \\
&\multicolumn{1}{c}{d$^{-1}$} & \multicolumn{1}{c}{mmag} &
\multicolumn{1}{c}{radians}  \\
& & \multicolumn{1}{c}{$\pm 0.001$} &\\

\midrule

$-\nu_1 + \nu_2  $ & $0.4087346  $ & $0.110  $ & $0.9693 \pm 0.0107 $ \\
$-2\nu_1 + 2\nu_2 $ & $0.8174692  $ & $0.007  $ & $1.4095 \pm 0.1584 $ \\
 $\nu_1 $ & $3.2699899 $ & $0.768  $ & $1.4283 \pm 0.0015 $ \\
$\nu_2 $ & $3.6787245 $ & $1.117  $ & $2.5033 \pm 0.0011 $ \\
$-\nu_1 + 2\nu_2 $ & $4.0874591  $ & $0.779  $ & $-0.8178 \pm 0.0015 $ \\
 $2\nu_1 $ & $6.5399798  $ & $0.013  $ & $0.0892 \pm 0.0921 $ \\
 $\nu_1 + \nu_2 $ & $6.9487144  $ & $0.103  $ & $-0.4452 \pm 0.0115 $ \\
$2\nu_2 $ & $7.3574490  $ & $0.886  $ & $-2.6611 \pm 0.0013 $ \\

\bottomrule

\end{tabular}
\label{table:545}
\end{table}

\subsection{The phases}

\begin{figure}
\centering	
\includegraphics[width=0.9\linewidth,angle=0]{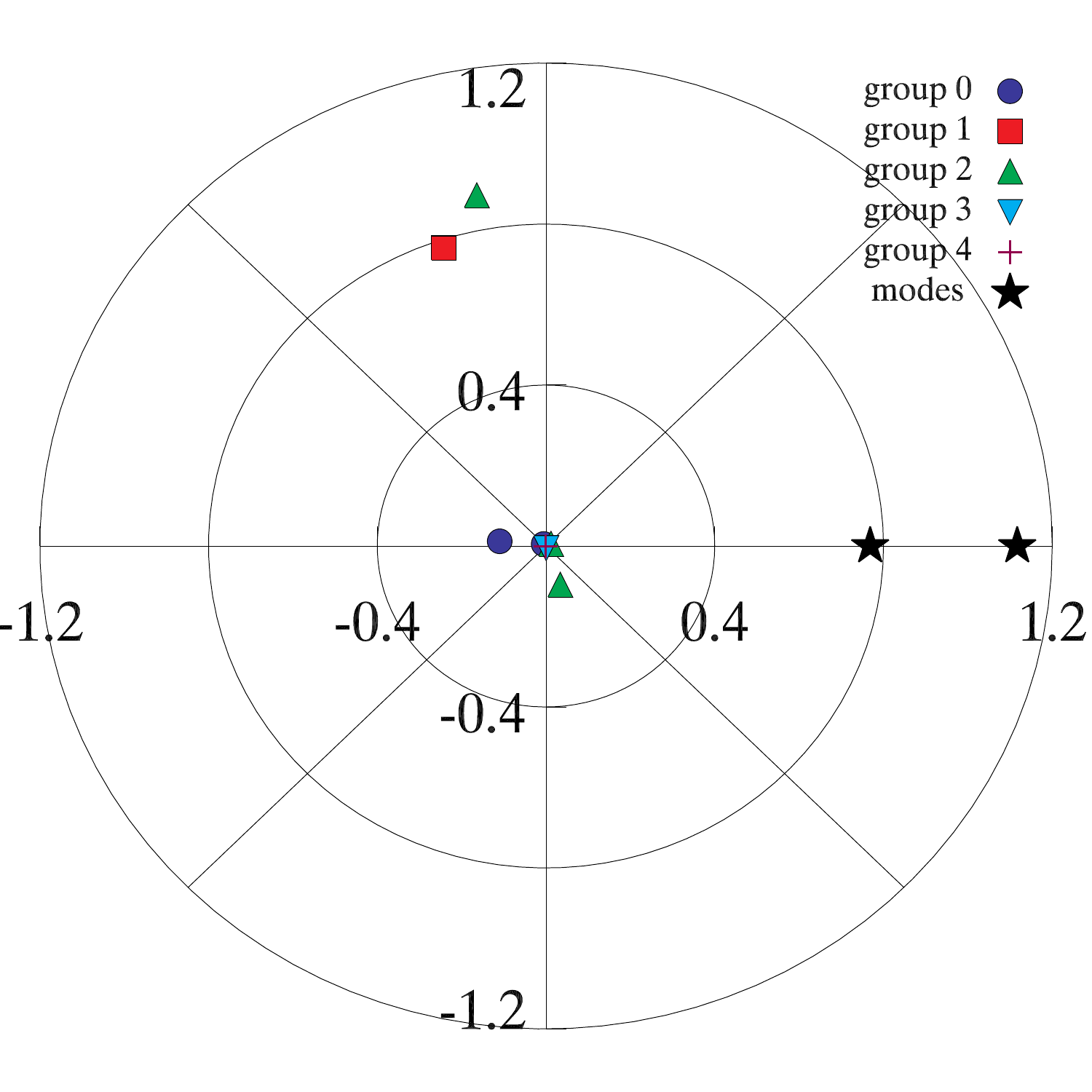}
\caption{The phasor plot for the SPB star KIC\,5450881.}
\label{fig:545_phasor}
\end{figure}

The light curve of KIC\,5450881 is not strongly asymmetric, hence the combination frequencies describing it are few in number and are not heavily biased to one side of the phasor diagram shown in Fig.\,\ref{fig:545_phasor}. Note that a large amplitude but a relative phase close to $\upi/2$ or $3\upi/2$ does not imply an asymmetric light curve; that is, a symmetric light curve could have many such peaks.

\section{B\MakeLowercase{e} stars}
\label{sec:bestars}

Several types of stars are given the spectral classification Be (B emission line) stars. There are pre-main-sequence Herbig Be stars (usually grouped with the Ae stars), there are binary mass transfer systems where circumstellar emission occurs, and there is a large class of stars that are rapidly rotating and have occasional outbursts where material is launched from the stellar surface into a disk, thus producing emission lines. Members of the latter group have been shown to be pulsating stars \citep{riviniusetal2003} and a model has been proposed by \citet{dylankeeetal2014} of how pulsation, coupled with rapid, but sub-critical rotation, can launch material into orbit. 

During the outbursts of pulsational Be stars the pulsation amplitude increases. Some B stars observed by {\it Kepler} show pulsational outbursts at long time intervals, and we consider those to be pulsational Be stars, even though we do not have spectroscopic observations to show the line emission that should accompany these pulsation outbursts. Future spectroscopic observations can test this. 

We show here one example for which the light curve and amplitude spectrum -- during and outside of pulsational outburst -- are strikingly similar to the frequency groups of SPB stars and also the $\gamma$\,Dor stars that we have presented in this paper. This similarity gives strong support to the hypothesis that the pulsational Be stars eject material when several, or many,  g~modes come into phase with each other. We suppose that those modes are prograde sectoral g modes where the pulsation velocity is primarily horizontal and in the direction of rotation, so that it adds to the subcritical rotation velocity to reach critical velocity and launch a circumstellar disk. 

In a review of pulsation in B stars and the prospects of asteroseismology, \citet{aerts2006} concluded: `It seems that pulsating Be stars are complicated analogues of the SPB stars $\ldots$'.  We concur: The single pulsating Be stars are rapidly rotating SPB stars. No physics other than pulsation need be conjectured to understand these stars.

\subsection{An example of a pulsating Be star: KIC\,11971405}
\label{sec:1197}

KIC\,11971405 (HD\,186567) is a relatively bright ($V = 9.2$) B8\,V star that shows g-mode pulsations typical of SPB stars. \citet{mcnamaraetal2012} listed it as a frequency group star, while \citet{pigulskietal2009} describe the star as aperiodic from ASAS ground-based data. We show here that it is an SPB star with frequency groups, and with occasional pulsation `outbursts' typical of pulsating Be stars. We hypothesise that it is a Be star, and that the pulsation outbursts are the result of modes with very closely spaced frequencies that only come into phase at long intervals to build up to an outburst. 

\begin{figure}
\centering	
\includegraphics[width=0.99\linewidth,angle=0]{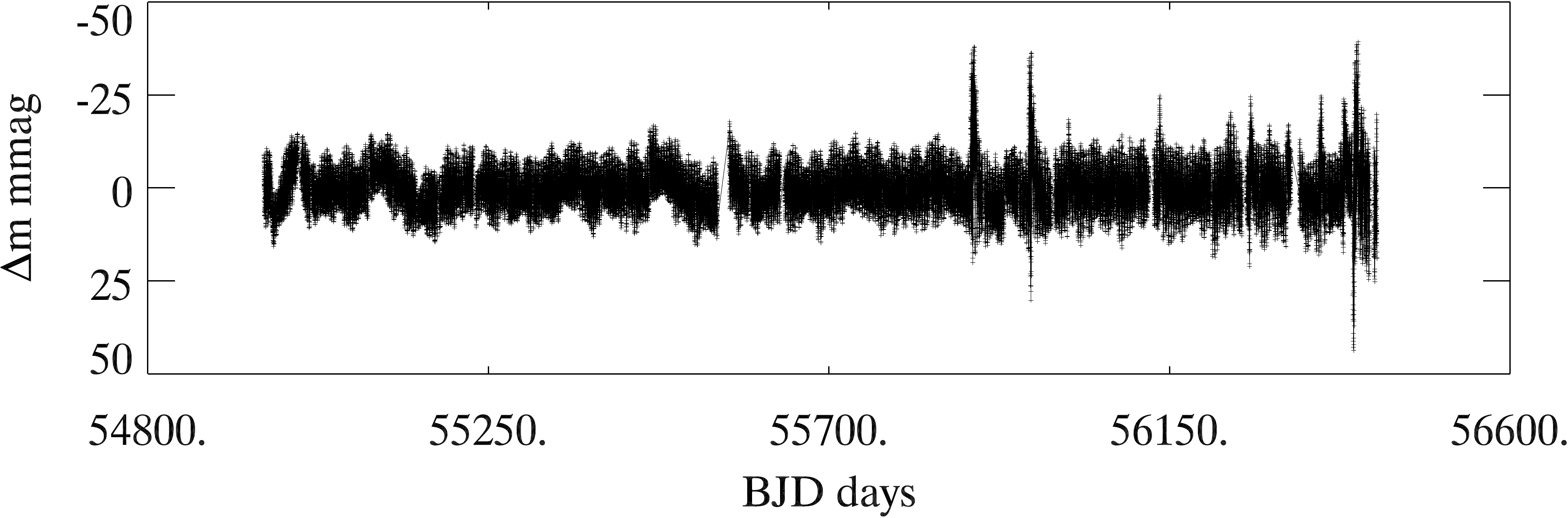}	
\includegraphics[width=0.99\linewidth,angle=0]{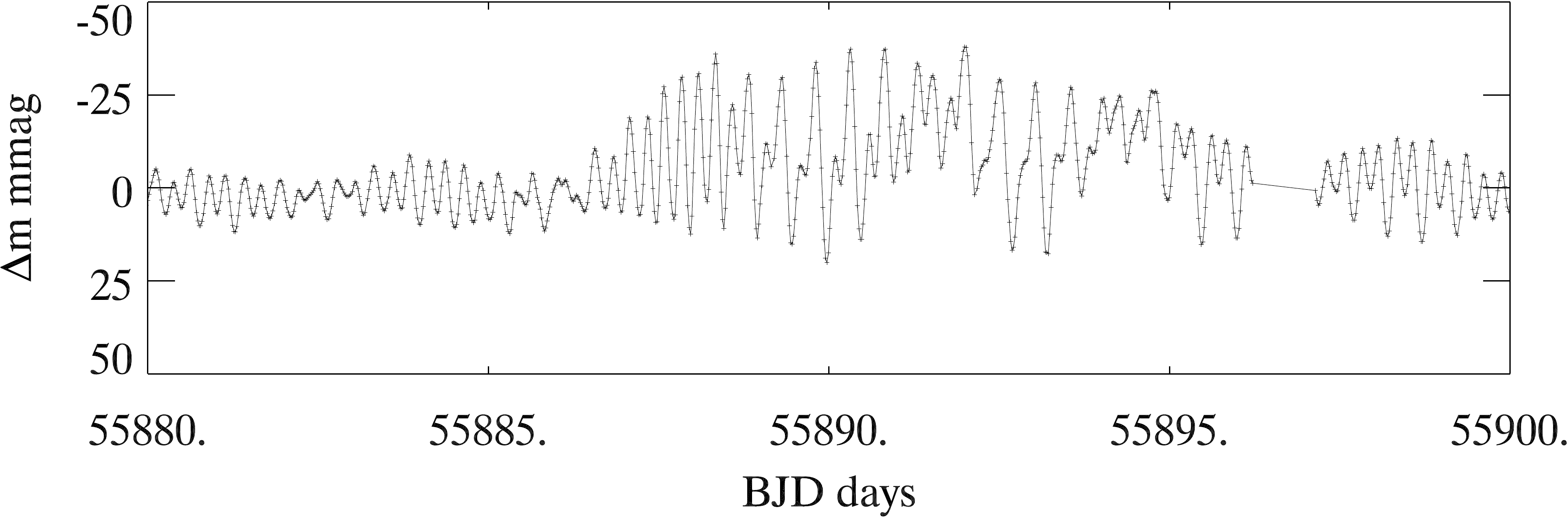}	
\includegraphics[width=0.99\linewidth,angle=0]{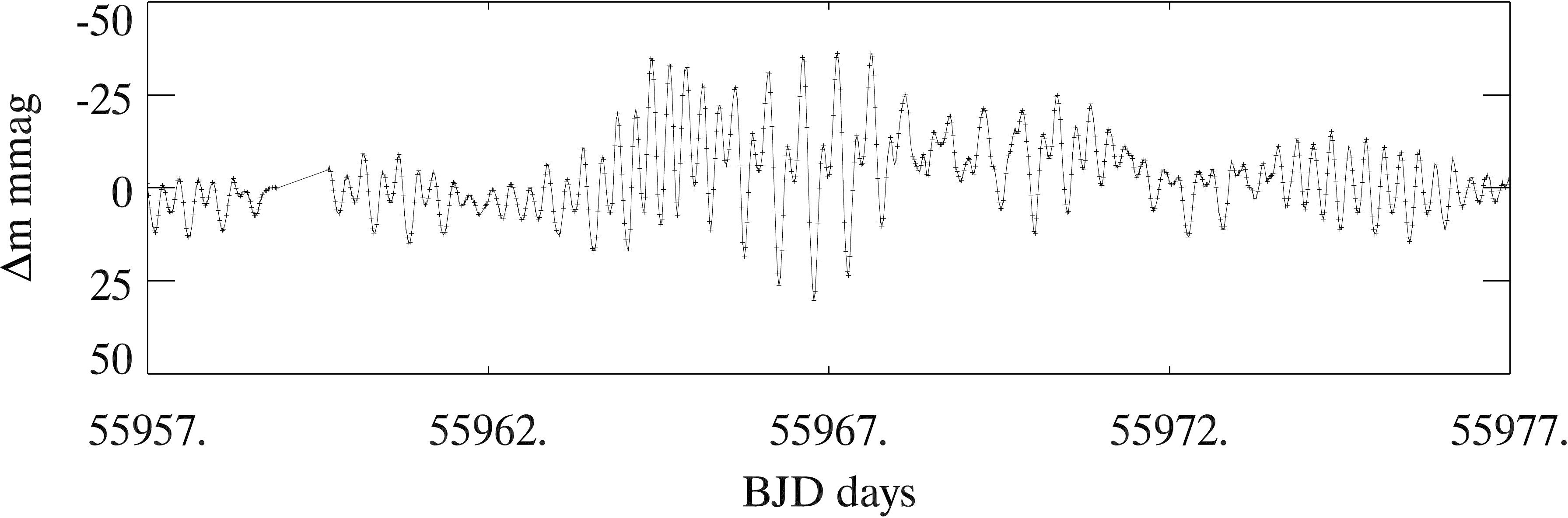}	
\includegraphics[width=0.99\linewidth,angle=0]{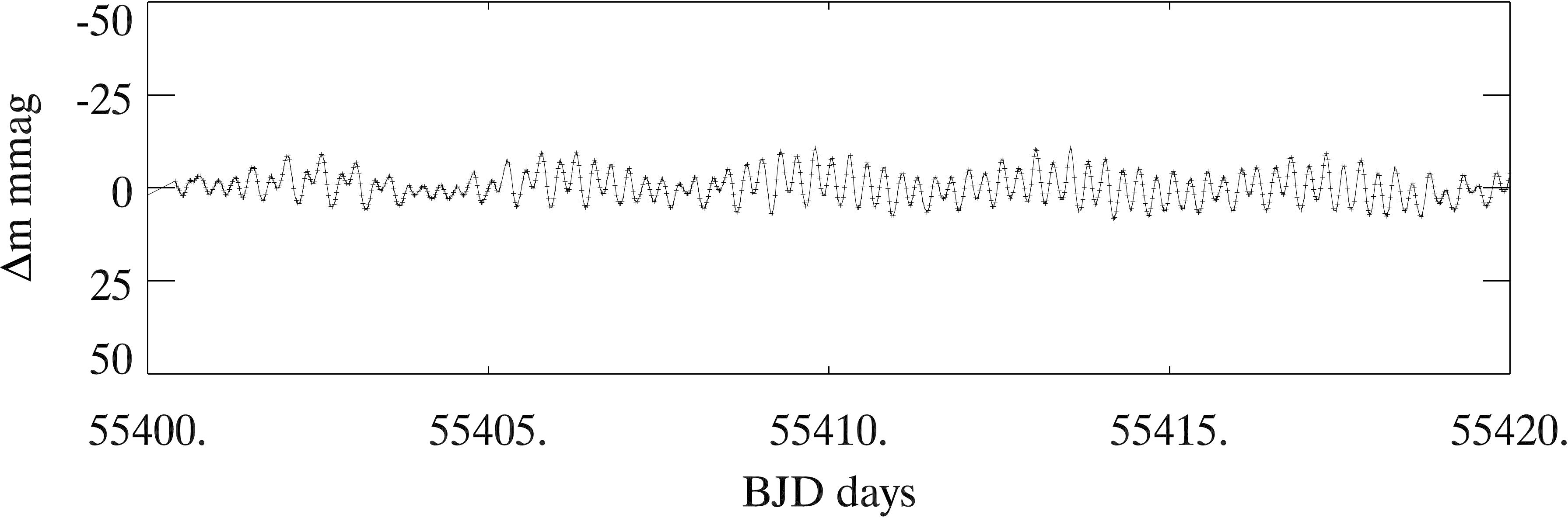}	
\includegraphics[width=0.99\linewidth,angle=0]{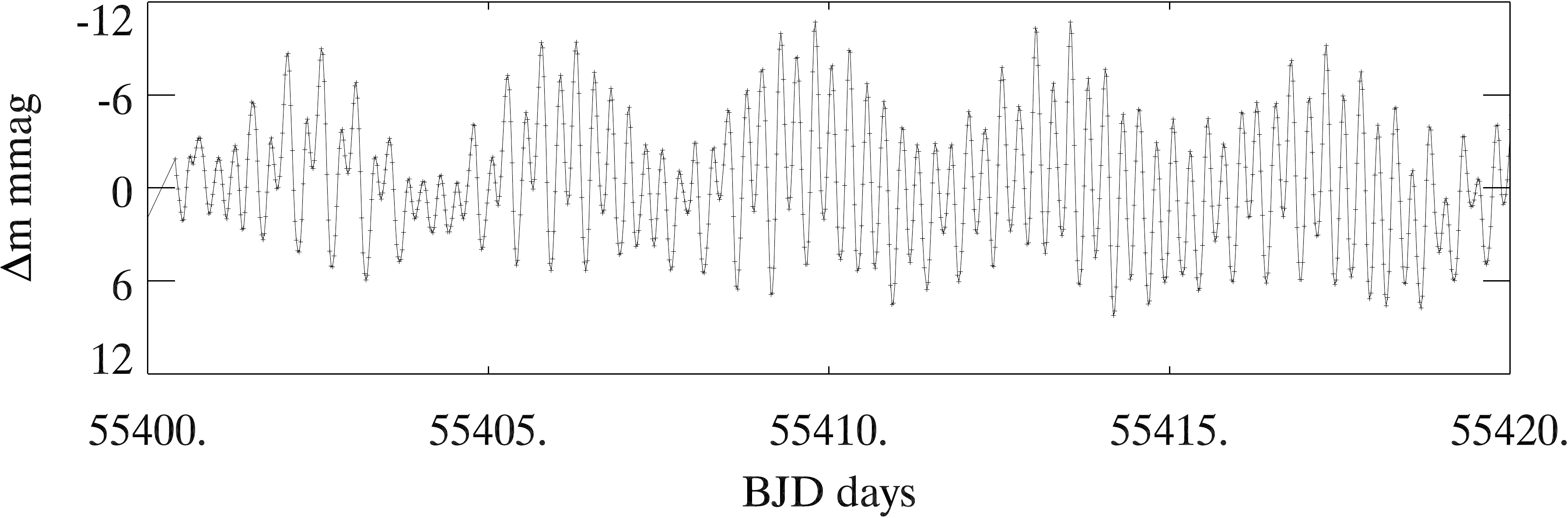}	
\caption{Top panel: The full {\it Kepler} Q0-17 light curve of KIC\,11971405. At this scale the pulsations are not visible, but the pulsation outbursts later in the mission can be seen. Second and third panels: 20-d sections of the light curve where two of the outbursts occur. The variations are typical symmetric or downward SPB pulsations. Fourth panel: a 20-d section outside of outburst at the same scale as the previous light curves. Bottom panel: the same as the fourth panel, but at expanded scale for visibility.}
\label{fig:1197_lc}
\end{figure}

Fig.\,\ref{fig:1197_lc} shows light curves for KIC\,11971405. The top panel shows the full {\it Kepler} Q0--17 data set. At this scale it is not possible to see the individual pulsations, but the pulsation outbursts later in the mission are evident. In the second and third panels, 20-d sections of the light curve show two of those outbursts in detail. The fourth panel shows a typical 20-d section of the star during quiescence at the same scale as the outburst light curves, and the bottom panel expands the ordinate scale for a more detailed look at the quiescent pulsational variations. KIC\,11971405 looks like a typical SPB star, except for the occasional outbursts.

\begin{figure}
\centering
\includegraphics[width=0.99\linewidth,angle=0]{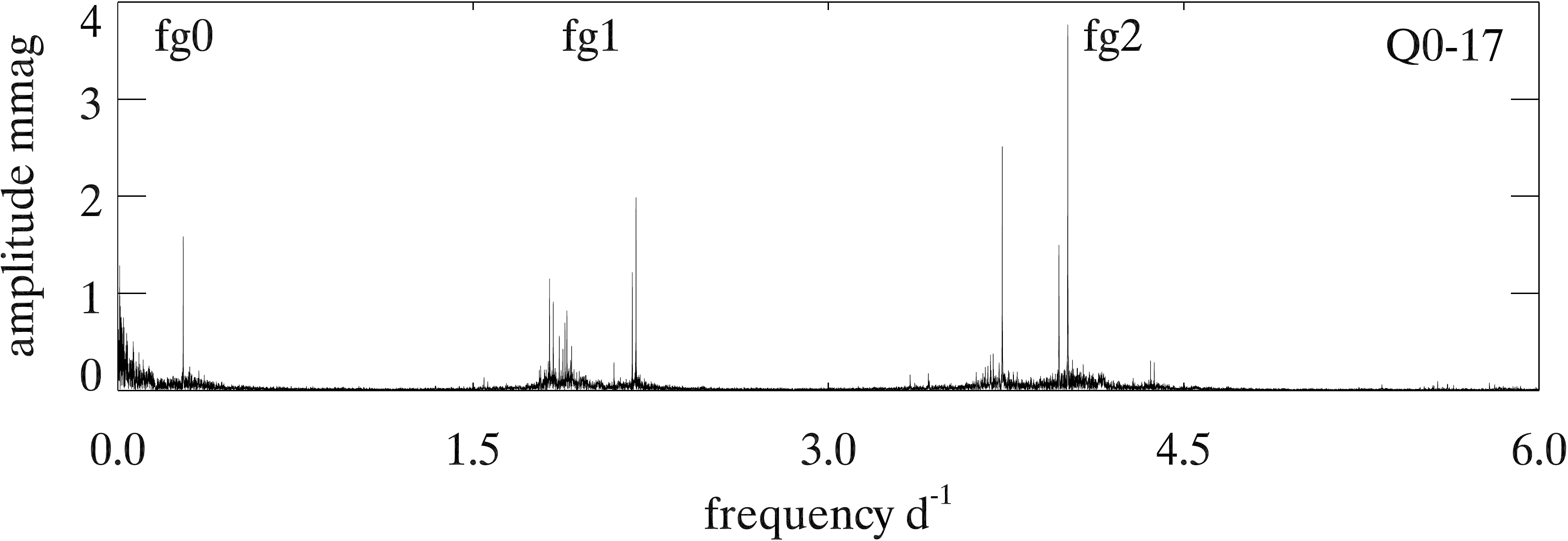}	
\includegraphics[width=0.99\linewidth,angle=0]{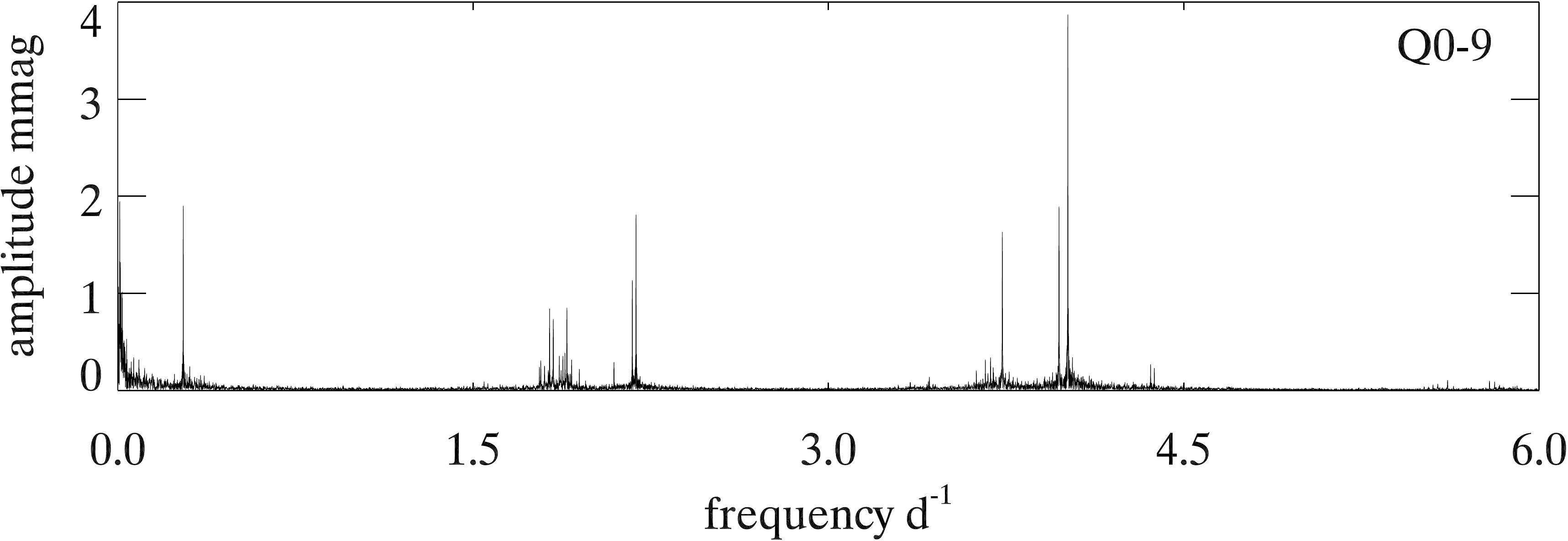}	
\includegraphics[width=0.99\linewidth,angle=0]{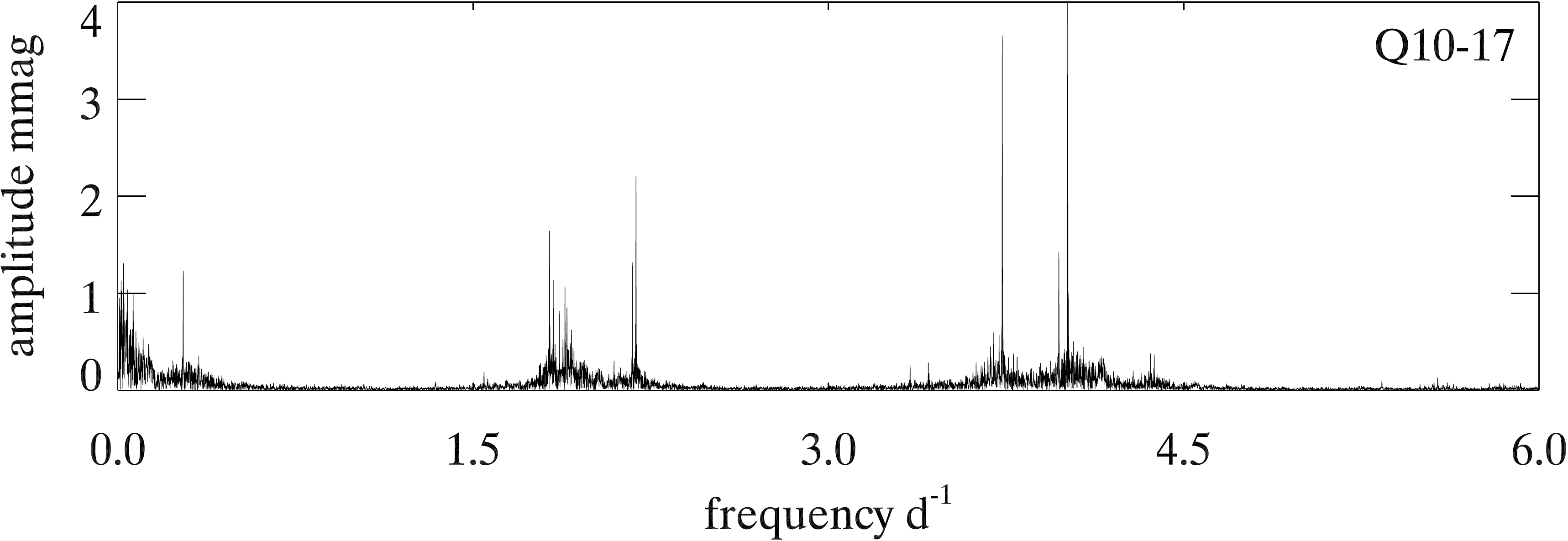}	
\caption{Top panel: An amplitude spectrum for KIC\,11971405 for the entire Q0--17 data set out to 6\,d$^{-1}$. There are some higher frequency groups at much lower amplitude that are not shown. Frequency groups fg0, fg1 and fg2 are marked. Although the highest amplitude peaks are in fg2, those are combinations of frequencies in fg1. Middle and bottom panels: The same as the top panel, but for independent data subsets Q0--9 and Q10--17. These show that the amplitudes of some of the peaks have changed, but the frequencies and the frequency patterns remain fixed.  }
\label{fig:1197_ft1}
\end{figure}

Fig.\,\ref{fig:1197_ft1} shows amplitude spectra for the entire Q0--17 data set (top) and for the independent subsets of the data Q0--9 and Q10--17 (middle and bottom). There are three main frequency groups, with further groups at much lower amplitude at higher frequency that are not displayed at this scale. This figure shows that the amplitudes of some of the peaks change between the independent data subsets, but the frequencies and their patterns remain the same. We hypothesise that this is caused by some pulsation mode frequencies that are so close in frequency that they are not resolved in the 4-yr time span of the {\it Kepler} data set. Those beat against each other slowly -- hence the long time span between pulsation outbursts -- and also come into phase at times with other pulsation modes. Outbursts arise when many modes come into phase to give the larger amplitudes seen in the light curves in the second and third panels of Fig.\,\ref{fig:1197_lc}.

A closer look at the amplitude spectrum for Q0--9, where there were no large outbursts, shows that the amplitude spectrum is dominated by combination frequencies. In particular, the highest amplitude peaks in fg2 are all simple combinations of frequencies in fg1. In Section\,\ref{sec:theory} we showed how it is possible for the combination frequencies to have higher amplitudes than the base frequencies. 
 
\begin{figure}
\centering	
\includegraphics[width=0.99\linewidth,angle=0]{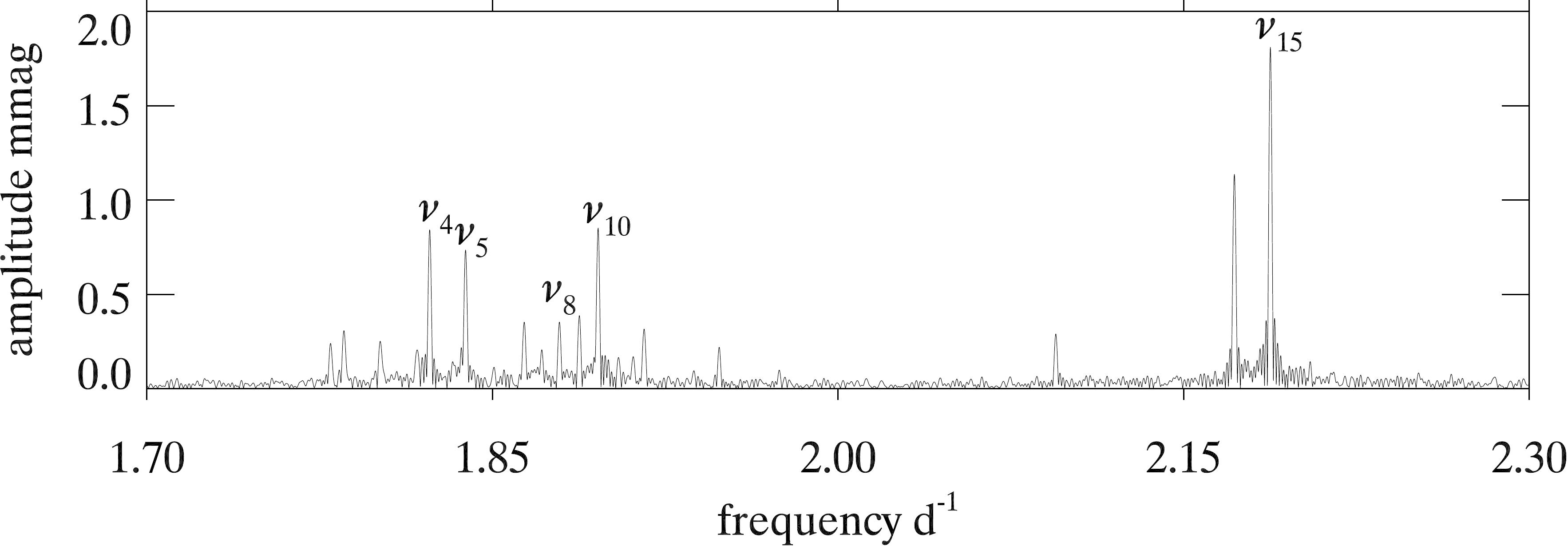}	
\includegraphics[width=0.99\linewidth,angle=0]{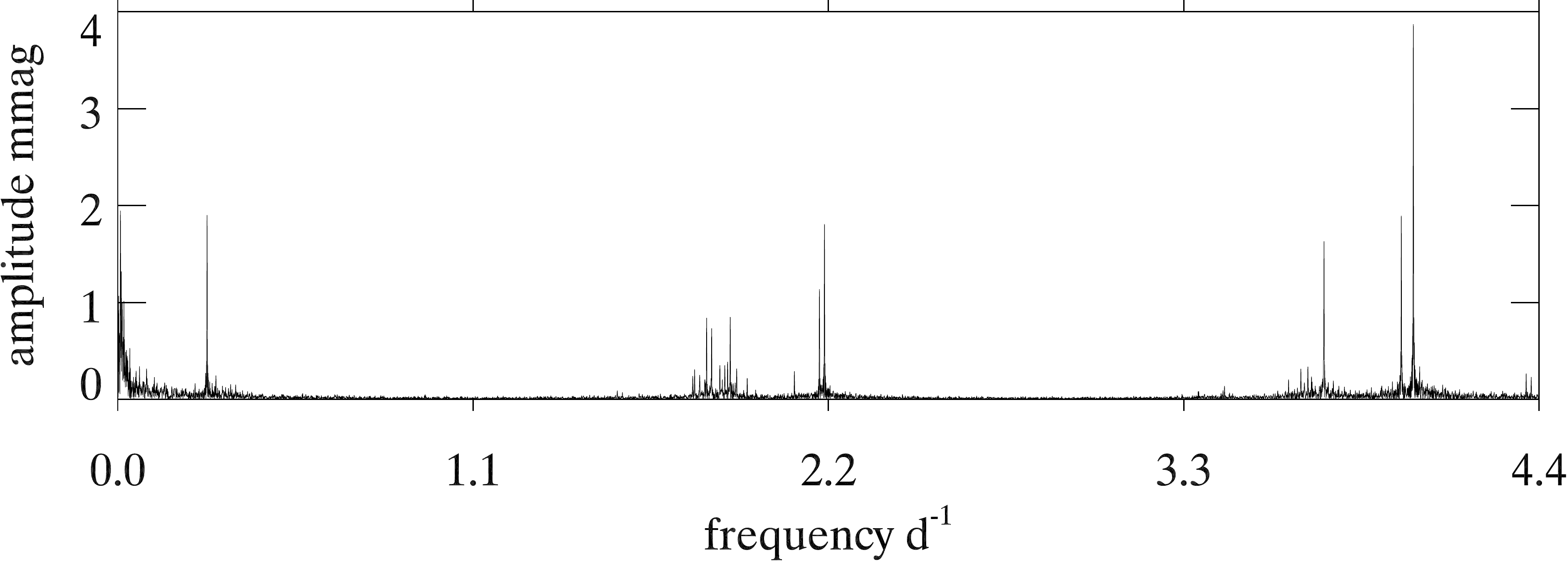}	
\includegraphics[width=0.99\linewidth,angle=0]{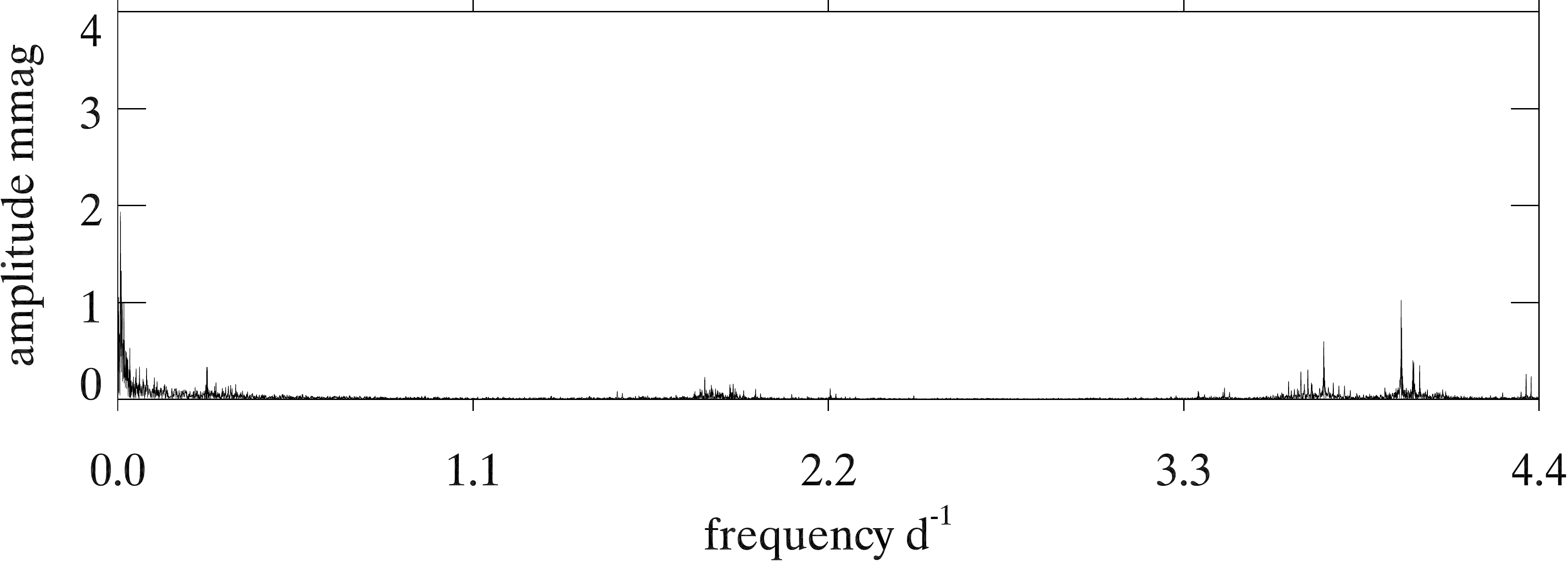}	
\caption{Top panel: An amplitude spectrum for KIC\,11971405 for the  Q0--9 data set  with some peaks in fg1 labelled. We use these as base frequencies to calculate combination frequencies and show that the high amplitude peaks in fg2 and fg0 are simple combinations of  frequencies in fg1. Middle panel: an amplitude spectrum for Q0--9. Bottom panel: An amplitude spectrum for Q0--9 after pre-whitening the 15 frequencies selected in fg1 and four combination frequencies that remove most of the variance in fg0 and fg2. }
\label{fig:1197_ft2}
\end{figure}

Fig.\,\ref{fig:1197_ft2} shows the amplitude spectrum for the Q0--9 data in the range of fg1 with a series of peaks identified. We selected from these our base frequencies for calculating the combination frequencies. Because there are so many frequencies in this group, we chose not to generate the thousands of combination terms that possibly arise among them, but instead selected just a few simple combinations to illustrate that the peaks in fg2, in particular, but also fg0, are combination frequencies. The frequency of the peak in fg0 is equal to the difference of the frequencies of the highest amplitude peaks in fg2, but those themselves are simple sum combination frequencies of base frequencies in fg1. 

We have thus shown that the SPB star KIC\,11971405, which shows pulsational outbursts, has a light curve and amplitude spectrum similar to those of other SPB stars and $\gamma$~Dor stars that show frequency groups. The principal difference is that the amplitudes of some of the peaks in the amplitude spectrum of KIC\,11971405 are variable over the 4-yr time span of the data set. We hypothesise that this is caused by pulsation modes with frequencies too closely spaced to be resolved in 4\,yr. This example and others in the {\it Kepler} data, support the hypothesis that the pulsational Be stars are rapidly rotating SPB stars.

\begin{table}
\centering
\caption[]{A least-squares fit of the 15 frequencies in fg1 of KIC\,11971405 plus four combination frequencies from four base frequencies in the fg1 set. The three highest peaks in fg2 and the peak in fg0 are thus shown to be combination frequencies. The residual amplitude at those frequencies is a consequence of the change in amplitude, probably because unresolved frequencies close to the base frequencies. The zero point of the time scale is ${\rm BJD} \,245\,5694.25$}
\begin{tabular}{clrr}

\toprule

\multicolumn{1}{c}{labels} &
\multicolumn{1}{c}{frequency} & \multicolumn{1}{c}{amplitude} &
\multicolumn{1}{c}{phase} \\
&\multicolumn{1}{c}{d$^{-1}$} & \multicolumn{1}{c}{mmag} &
\multicolumn{1}{c}{radians}  \\
& & \multicolumn{1}{c}{$\pm 0.004$} &\\

\midrule
\multicolumn{4}{c}{fg0}\\
\midrule

$\nu_4+\nu_{15} - \nu_5 - \nu_{10} $ & $ 0.276260 $ & $ 1.858 $ & $ 0.809 \pm 0.002 $ \\

\midrule
\multicolumn{4}{c}{fg1}\\
\midrule

$\nu_1 $ & $ 1.779808 $ & $ 0.236 $ & $ -2.532 \pm 0.017 $ \\
$\nu_2 $ & $ 1.785546 $ & $ 0.318 $ & $ 0.200 \pm 0.013 $ \\
$\nu_3 $ & $ 1.801339 $ & $ 0.207 $ & $ -0.644 \pm 0.019 $ \\
$\nu_4 $ & $ 1.822719 $ & $ 0.859 $ & $ 1.824 \pm 0.005 $ \\
$\nu_5 $ & $ 1.838304 $ & $ 0.736 $ & $ 0.303 \pm 0.005 $ \\
$\nu_6 $ & $ 1.863744 $ & $ 0.383 $ & $ 0.451 \pm 0.010 $ \\
$\nu_7 $ & $ 1.871133 $ & $ 0.163 $ & $ -2.767 \pm 0.025 $ \\
$\nu_8 $ & $ 1.879095 $ & $ 0.362 $ & $ 0.541 \pm 0.011 $ \\
$\nu_9 $ & $ 1.887687 $ & $ 0.430 $ & $ 2.551 \pm 0.009 $ \\
$\nu_{10} $ & $ 1.895767 $ & $ 0.850 $ & $ 3.116 \pm 0.005 $ \\
$\nu_{11} $ & $ 1.915925 $ & $ 0.309 $ & $ -0.289 \pm 0.013 $ \\
$\nu_{12} $ & $ 1.948357 $ & $ 0.227 $ & $ 2.728 \pm 0.018 $ \\
$\nu_{13} $ & $ 2.094494 $ & $ 0.287 $ & $ 2.116 \pm 0.014 $ \\
$\nu_{14} $ & $ 2.171925 $ & $ 1.166 $ & $ 0.071 \pm 0.003 $ \\
$\nu_{15} $ & $ 2.187612 $ & $ 1.826 $ & $ -1.173 \pm 0.002 $ \\

\midrule
\multicolumn{4}{c}{fg2}\\
\midrule

$\nu_5+\nu_{10} $ & $ 3.734071 $ & $ 1.571 $ & $ -2.086 \pm 0.003 $ \\
$\nu_8+\nu_{13} $ & $ 3.973589 $ & $ 1.412 $ & $ -0.441 \pm 0.003 $ \\
$\nu_4 + \nu_{15} $ & $ 4.010331 $ & $ 3.867 $ & $ 1.032 \pm 0.001 $ \\

\bottomrule

\end{tabular}
\label{table:1197}
\end{table}

\section{Discussion and conclusions}

The frequency spectra of the g-mode pulsators of the main-sequence, the $\gamma$~Dor stars and the SPB stars, show a wide variety of complexity. Some stars show series of consecutive radial overtone g~mode frequencies, and it is those that are the most interest asteroseismically, since they allow inference about the core conditions in the stars (e.g., \citealt{vanreethetal2015}; \citealt{saioetal2015}; \citealt{kurtzetal2014}; \citealt{beddingetal2014}; \citealt{bouabidetal2013}). Other stars show frequency groups that have not been understood, or have been only partially understood, until now. We have shown in this paper that the frequency groups found in the $\gamma$~Dor stars and SPB stars are often dominated by combination frequencies of only a few base frequencies, and we interpret those base frequencies as arising from a small number of g~modes. 

We have shown that the shapes of the light curves, which have previously been classified by visual inspection into `symmetric' and `asymmetric' are a consequence of the same physics. It is the phases of the combination frequencies that describe the visual appearance of the light curves. We have extended this to show that `downward' light curves, which some investigators have conjectured are caused partially or completely by stellar spots, are also explained by a small number of g~mode pulsations and their combination frequencies. We further showed that SPB stars with pulsational outbursts show frequency groups and combination frequencies that are the same in character to other SPB stars, supporting to the conclusion that pulsational Be stars are rapidly rotating SPB stars.

The only physics necessary to understand the light curves of the $\gamma$~Dor stars, the SPB stars and the pulsating Be stars is g-mode pulsation. For many years, there has been conjecture that spots are a source of the light curve shapes in these stars. A particular driver for that conjecture is the shape of the light curves of the stars that show `downward' light curves. We have shown that pulsation alone accounts for the light curve shapes.

For the $\gamma$~Dor and SPB stars presented in this paper, the base frequencies and the combination frequencies are unchanged to high precision over the 4-yr time span of the data. No spot model can account for this. We know what stable spots look like on upper main-sequence stars from the chemically peculiar magnetic Bp, Ap and Fp stars. Those show rotational light curves that are non-sinusoidal, hence can have many harmonics of the rotational frequency. But there are no combination frequencies generated. In the frequency groups of the stars studied in this paper, the combination frequencies dominate much of the amplitude spectrum and harmonics of the base frequencies have low amplitudes; this is not the signature of spots. 

For asteroseismic studies of SPB stars and $\gamma$~Dor stars, combination frequencies have been considered a nuisance to be detected and discarded in the acquisition of pulsation mode frequencies for asteroseismology. We now see that in some cases they can comprise the major components of the amplitude spectra. Indeed, for asteroseismology they may still be modelled and removed. But it may be possible to use the combination frequencies for asteroseismic study themselves, as has been done for white dwarf stars for many years and as is being attempted for p~mode pulsations in $\delta$~Sct stars. The large mode energies implied by the visibility of surface g~modes promise the discovery of interesting new physical information about the cores of these main-sequence stars.

\section*{acknowledgements}

DWK thanks the Japan Society for the Promotion of Science (JSPS) for an Invitation Fellowship, during which much of this work was carried out at the University of Tokyo. We thank Prof. Hideyuki Saio and Dr Masao Takata for useful discussions. This research was supported by the Australian Research Council. Funding for the Stellar Astrophysics Centre is provided by the Danish National Research Foundation (grant agreement no.: DNRF106). The research is supported by the ASTERISK project (ASTERoseismic Investigations with SONG and Kepler) funded by the European Research Council (grant agreement no.: 267864).

\bibliography{arxiv_fgstars}

\end{document}